\documentclass[a4paper,nofootinbib,preprintnumbers,twocolumn,preprintnumbers,floatfix,superscriptaddress,pra,showpacs]{revtex4-1}

\usepackage{graphicx}
\usepackage{amssymb}
\usepackage{amsmath}
\usepackage{epstopdf}
\usepackage{graphics}
\usepackage[caption=false]{subfig}
\usepackage{float}
\usepackage{color}
\usepackage[colorlinks=true,urlcolor=blue]{hyperref}  
\usepackage{bbold}

\allowdisplaybreaks[1]

\newcommand{\bra}[1]{\langle #1|}					
\newcommand{\ket}[1]{|#1\rangle}					

\newcommand{\abs}[1]{\left| #1 \right|} 
\newcommand{\avg}[1]{\langle #1 \rangle} 

\newcommand{\figref}[1]{Fig.~\ref{#1}}

\begin{document}

\title{One- and two-axis squeezing of atomic ensembles in optical cavities}

\author{J. Borregaard }
\affiliation{The Niels Bohr Institute, University of Copenhagen, Blegdamsvej 17, 2100 Copenhagen \O, Denmark}
\affiliation{Department of Physics, Harvard University, Cambridge, MA 02138, USA}
\affiliation{QMATH, Department of Mathematical Sciences, University of Copenhagen, Universitetsparken 5, 2100 Copenhagen \O, Denmark}
\author{E. J. Davis}
\affiliation{Department of Physics, Stanford University, Stanford, CA 94305, USA}
\author{G. S. Bentsen}
\affiliation{Department of Physics, Stanford University, Stanford, CA 94305, USA}
\author{M. H. Schleier-Smith}
\affiliation{Department of Physics, Stanford University, Stanford, CA 94305, USA}
\author{A. S. S\o rensen}
\affiliation{The Niels Bohr Institute, University of Copenhagen, Blegdamsvej 17, 2100 Copenhagen \O, Denmark}

\date{\today}

\begin{abstract}
The strong light-matter coupling attainable in optical cavities enables the generation of highly squeezed states of atomic ensembles.  It was shown in [\href{http://link.aps.org/doi/10.1103/PhysRevA.66.022314}{Phys. Rev. A \textbf{66}, 022314 (2002)}] how an effective one-axis twisting Hamiltonian can be realized in a cavity setup. Here, we extend this work and show how an effective two-axis twisting Hamiltonian can be realized in a similar cavity setup. We compare the two schemes in order to characterize their advantages. In the absence of decoherence, the two-axis Hamiltonian leads to more squeezing than the one-axis Hamiltonian. If limited by decoherence from spontaneous emission and cavity decay, we find roughly the same level of squeezing for the two schemes scaling as $\sqrt{NC}$  where $C$ is the single atom cooperativity and $N$ is the total number of atoms. When compared to an ideal squeezing operation, we find that for specific initial states, a dissipative version of the one-axis scheme attains higher fidelity than the unitary one-axis scheme or the two-axis scheme. However, the unitary one-axis and two-axis schemes perform better for general initial states. 
\end{abstract}


\maketitle

\section{Introduction}
Spin squeezed states of atomic ensembles have many applications as resources for quantum enhanced metrology~\cite{wineland2,andre,johannes1,leroux2010,Hosten2}, continuous variable quantum information processing~\cite{rev123}, and multipartite entanglement~\cite{anders2,korbicz,briegel1}. Various methods for generating spin squeezed states in atomic ensembles have been proposed~\cite{thorne,anders1,leroux2012,torre,monika5} and realized experimentally~\cite{eugene1,eugene2,muessel,monika1,leroux2010,vasilakis,bohnet}. In particular, cavity-based schemes where the light-matter interaction is enhanced by placing the atoms in an optical cavity have have shown impressive results and have realized highly squeezed states~\cite{leroux2012,Hosten2,bohnet}. To take full advantage of these experimental advances and to ensure a continued increase in their capabilities, it is important to  determine the ideal operation conditions and the squeezing attainable with such cavity based approaches. 

A commonly used measure for the degree of squeezing in an ensemble is the possible gain in precision by using the squeezed state for interferometry. Wineland \emph{et al.}~\cite{wineland1} showed that this can be quantified by
\begin{equation} \label{eq:xi2}
\xi^2=\underset{\theta}{\min}\left(\frac{N\left(\avg{\hat{J}_{\theta}^2}-\avg{\hat{J}_{\theta}}^2\right)}{\avg{\hat{J}_z}^2}\right),
\end{equation}
where $\avg{\hat{J}_z}\approx N/2$ is the mean spin and $\hat{J}_{\theta}=\cos(\theta)\hat{J}_x+\sin(\theta)\hat{J}_y$. Here, $\hat{J}_{x,y,z}$ are the collective spin operators defined in the usual manner~\cite{wineland1}. For $\xi^2<1$ a gain in interferometric precision is possible compared to using a coherent spin state. 

In general, cavity based schemes are known to exhibit a $1/\sqrt{NC}$ scaling of $\xi^2$ when limited by dissipation. Here, $C$ is the single atom cooperativity (defined below) and $N$ is the total number of atoms. This scaling is obtained as a tradeoff between the competing processes of the coherent evolution causing squeezing and the dissipative processes of spontaneous emission and cavity decay~\cite{torre,monika5,anders1}.

The squeezing parameter $\xi^2$ is, however, not a complete characterization of the dynamics. The precise figure of merit will depend on the application for which the squeezing operation is used, and so may the optimal method of squeezing.  For example, if the objective is to prepare a specific squeezed state for metrology, dissipative schemes~\cite{torre} where the system is driven into a squeezed dark state may be beneficial.  However, in continuous variable quantum information processing applications~\cite{rev123} where the objective is to implement a squeezing operation on a generic input state, coherent schemes~\cite{anders1,takeuchi2005spin,leroux2012,zhang2015,trail2010} may be advantageous.



A demonstrated approach to coherent spin squeezing is to implement a one-axis twisting Hamiltonian \cite{ueda}:
\begin{equation} \label{eq:hamil1}
\hat{H}_{1-\text{axis}}=\alpha\hat{J}_{\theta}^2.
\end{equation}
This non-linear Hamiltonian has already been realized for atoms in optical cavities \cite{anders1,ueda,monika5,hosten}, and in several other physical systems \cite{gross2010nonlinear,riedel2010atom,bohnet2016quantum}.  Theoretically, squeezing can also be induced by the two-axis countertwisting Hamiltonian
\begin{equation} \label{eq:hamil2}
\hat{H}_{2\text{-axis}}=\alpha\left(\hat{J}_{\theta}^2-\hat{J}_{\theta+\frac{\pi}{2}}^2\right),
\end{equation}
which may offer advantages over one-axis twisting.  In the absence of decoherence, $\hat{H}_{2\text{-axis}}$ leads to Heisenberg limited squeezing, $\xi^{2}\sim1/N$, which is the fundamental limit~\cite{ueda}. This is in contrast to the one-axis twisting Hamiltonian~\eqref{eq:hamil1}, which has a theoretical limit of $\xi^{2}\sim1/N^{\frac{2}{3}}$ arising from the curvature of the Bloch sphere~\cite{ueda,davis1}.  Furthermore, the two-axis Hamiltonian squeezes exponentially in time while the one-axis Hamiltonian squeezes only polynomially~\cite{opa2015}. This has motivated efforts to realize two-axis Hamiltonians in various settings~\cite{liu2011,kruse2016}.

\begin{figure} [h]
\centering
\includegraphics[width=0.4\textwidth]{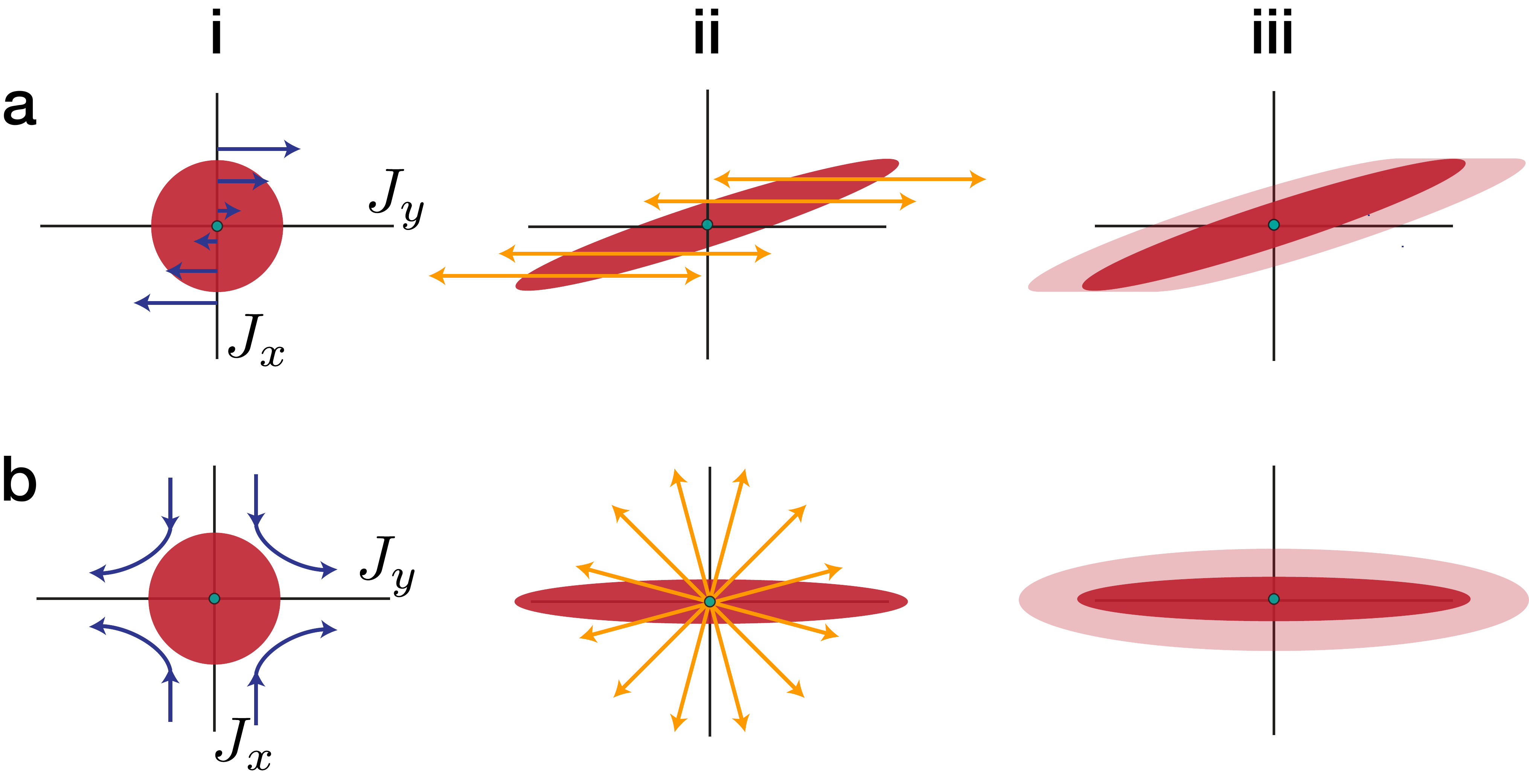}
\caption{Schematic illustration of unitary evolution and added noise due to collective dissipation in both the one-axis twisting and two-axis countertwisting  schemes. Row (a) shows one-axis twisting: in (i) the coherent spin state deforms under $\hat{H}_{1-\text{axis}}=\alpha\hat{J}_x^2$ as indicated by the blue flow lines, resulting in a squeezed state (ii). In practice, collective dissipation broadens the state along $\hat{J}_y$ (orange flow lines in (ii)), resulting in added noise (iii) which is mostly in the anti-squeezed quadrature. Row (b) shows the two-axis countertwisting evolution: in (i) the coherent spin state deforms under $\hat{H}_{2-\text{axis}}$ as indicated by the blue flow lines, resulting in a squeezed state (ii). Collective dissipation broadens the state in all directions in the $xy$-plane (orange flow lines in (ii)), resulting in added noise (iii) which affects the squeezed and anti-squeezed quadrature in a similar manner.}
\label{fig:figure0}
\end{figure}

In this article, we extend the cavity-based one-axis twisting scheme of Ref.~\cite{anders1} to show how an effective two-axis twisting Hamiltonian can be engineered.  For atoms strongly coupled to the cavity such that dissipation can be neglected, the two-axis scheme creates stronger squeezing than the one axis scheme. However, for weakly coupled atoms the situation is different. We find that when limited by decoherence, $\xi^2$ scales as $1/\sqrt{NC}$ for both the one- and two-axis schemes and the two schemes exhibit similar amounts of squeezing. We find that this is because the collective decay adds more noise to the squeezed quadrature in the two-axis scheme than the one-axis scheme, as shown qualitatively in \figref{fig:figure0}. For quantum information processing, not only the amount of squeezing but also the purity of the squeezing operation matters~\cite{rev123}. We therefore also compare the performance of both schemes to an ideal squeezing operation. We find that also in this case, the one-axis scheme performs similar to or better than the two-axis scheme when limited by decoherence. 




In the one-axis twisting scheme of Ref.~\cite{anders1}, a collection of atoms is placed in a cavity such that two ground states are both coupled off-resonantly through the cavity field to an excited state (\figref{fig:figure1}). By illuminating the atoms with bichromatic light, pairwise exchange between the ground states can be realized, resulting in the quadratic Hamiltonian $\hat{H}_{1-\text{axis}}=\alpha\hat{J}_{\theta}^2$.  Below, we first show that by adding a second bichromatic laser to the setup of Ref.~\cite{anders1}, the effective dynamics can be described by a two-axis twisting Hamiltonian.  We then proceed by analyzing and comparing the squeezing properties of both the original one-axis scheme and the modified two-axis scheme, including the effects of dissipation. Finally, we elaborate on the requirements for the validity of the effective dynamics considered.

\section{Effective dynamics}
\begin{figure} [h]
\centering
\includegraphics[width=0.4\textwidth]{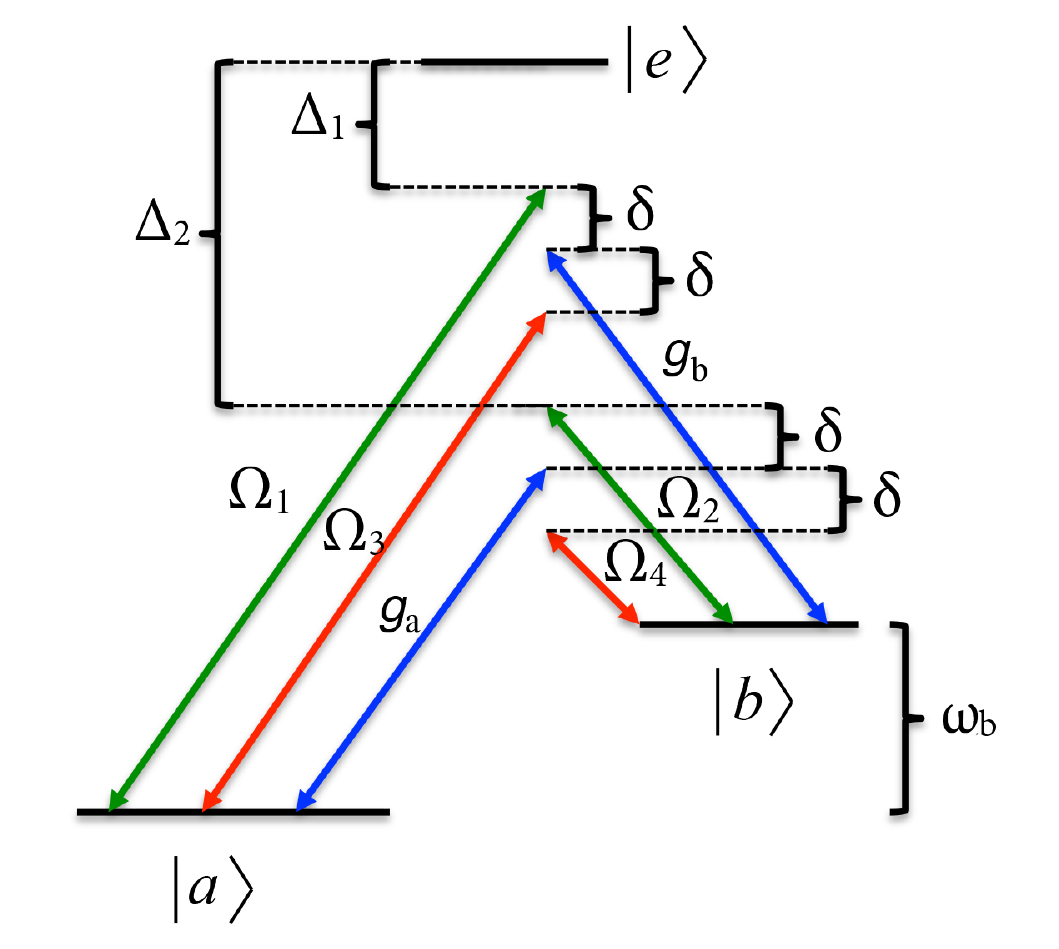}
\caption{Atomic level structure illustrating the laser fields ($\Omega$) and cavity couplings ($g$). The atomic detunings $\Delta_{1,2}$ and the two-photon detuning $\delta$ are also shown. }
\label{fig:figure1}
\end{figure}

We assume that the atoms have two stable ground states $\ket{a}$ and $\ket{b}$ and an excited level $\ket{e}$. The ground states are coupled to the excited level through four laser couplings and two cavity couplings with coupling constants $g_{\text{a}}$ and $g_{\text{b}}$ as shown in \figref{fig:figure1}. 
In a suitable rotating frame, the Hamiltonian describing the system is
\begin{eqnarray}
\hat{H}&=&\sum_{k=1}^{N}\left(\frac{\Omega_1}{2}e^{i\Delta_1t}+\frac{\Omega_3}{2}e^{i\Delta_3t}+g_{\text{a}}e^{i(\Delta_2+\delta)t}\hat{c}\right)\ket{e}_k\bra{a} \nonumber \\
&&+\left(\frac{\Omega_2}{2}e^{i\Delta_2t}+\frac{\Omega_4}{2}e^{i\Delta_4t}+g_{\text{b}}e^{i(\Delta_1+\delta)t}\hat{c}\right)\ket{e}_k\bra{b} \nonumber \\
&&+\text{H. c.},
\end{eqnarray}  
where H.c is the Hermitian conjugate. The $N$ atoms are labelled by the subscript $k$ and we have defined the detunings $\Delta_1=\omega_{\text{e}}-\omega_{\text{L}1}$, $\Delta_2=\Delta_1-\omega_{\text{b}}$, $\Delta_3=\Delta_1+2\delta$, $\Delta_4=\Delta_2+2\delta$, and $\delta=\omega_{\text{L}1}-\omega_{\text{b}}-\omega_{\text{cav}}$. Here, $\omega_{\text{e}}$ ($\omega_{\text{b}})$ is the transition frequency between level $\ket{a}$ and $\ket{e}$ ($\ket{b}$), $\omega_{\text{L}x}$ is the frequency of laser $x$, and $\omega_{\text{cav}}$ is the cavity resonance frequency. The four laser couplings are denoted $\Omega_{1-4}$ and $g_{\text{a}}$ ($g_{\text{b}}$) is the cavity coupling of level $\ket{a}$ ($\ket{b}$). We have assumed the upper of the two lasers addressing different transitions to differ by twice the ground state splitting $\omega_{\text{L}1}-\omega_{\text{L}2}=2\omega_{\text{b}}$ and similarly the lower two fields differ by the same amount $\omega_{\text{L}3}-\omega_{\text{L}4}=2\omega_{\text{b}}$. Furthermore, we have assumed that the laser fields addressing the same transitions differ in frequency by $2\delta$ so that $\omega_{\text{L}1}-\omega_{\text{L}3}=\omega_{\text{L}2}-\omega_{\text{L}4}=2\delta$. The decay of state $\ket{e}_{k}$ is assumed to be described by the Lindblad operators $\hat{L}_{x}^{(k)}=\sqrt{\gamma_{x}}\ket{x}_{k}\bra{e}$, where $\gamma_{x}$ is the decay rate into state $\ket{x}$ and $x\in\{a,b,o\}$. The state $\ket{o}$ represents all other  ground states than $\ket{a}$ and $\ket{b}$. The total decay rate of the excited state is $\Gamma=\gamma_{a}+\gamma_{b}+\gamma_{o}$. The decay of the cavity field is assumed to be described by the Lindblad operator $\hat{L}_{c}=\sqrt{\kappa}\hat{c}$, where $\kappa$ is the intensity decay rate of the cavity and $\hat{c}$ is the annihilation operator of the cavity field. We assume that both ground states are coupled to the excited state through the same cavity field.  

The basic mechanism behind the scheme can be understood from considering the various transitions mediated by the laser and cavity fields. Assuming large detunings, the couplings from laser 1 and 2 allows a two-photon resonant transitions of the form $\ket{aa}\to\ket{bb}$ ($\ket{bb}\to\ket{aa}$). Here, an atom in state $\ket{a}$ ($\ket{b}$) absorbs a photon from laser 1 (2) and emits a cavity photon that is absorbed by another atom in state $\ket{a}$ ($\ket{b}$), which then emits into laser 2 (1) resulting in the simultaneous transfer of two atoms from $\ket{a}$ to $\ket{b}$ ($\ket{b}$ to $\ket{a}$). Since laser 1 is detuned by $\delta$ and laser 2 by $-\delta$, processes involving only a single atom are off resonant and will be suppressed. In the two atom process, however, the two detunings cancel, making the total two atom process $\ket{aa}\to\ket{bb}$ ($\ket{bb}\to\ket{aa}$) resonant. The resulting dynamics can thus be described by a term $\hat{J}_+^2$ ($\hat{J}_-^2$) in an effective Hamiltonian for the ground states where $\hat{J}_{+}=\sum_{k}\ket{a}_{k}\bra{b}$ and $\hat{J}_{-}=\hat{J}_{+}^{\dagger}$. Other resonant processes are transitions of the form $\ket{ab}\to\ket{ba}$ ($\ket{ba}\to\ket{ab}$) where an atom in state $\ket{a}$ ($\ket{b}$) absorbs a photon from laser 1 (2) and emits a cavity photon that is absorbed by an atom in state $\ket{b}$ ($\ket{a}$), which then emits into laser 1 (2). These processes are described by a term $\hat{J}_{-}\hat{J}_+$ ($\hat{J}_{+}\hat{J}_-$) in the Hamiltonian. As a consequence, the effective Hamiltonian describing the evolution due to laser 1 and 2 is 
\begin{eqnarray}
\hat{H}_{\text{eff}}&\sim&\frac{\abs{\Omega_1}^2\abs{g_{\text{b}}}^2}{4\Delta_1^2\delta}\hat{J}_{+}\hat{J}_{-}+\frac{\abs{\Omega_2}^2\abs{g_{\text{a}}}^2}{4\Delta_2^2\delta}\hat{J}_{-}\hat{J}_{+}\nonumber \\
&&+\frac{\Omega_1^{*}g_{\text{b}}g_{\text{a}}^{*}\Omega_2}{4\Delta_1\Delta_2\delta}\hat{J}_{+}^2+\frac{\Omega_2^{*}g_{\text{a}}g_{\text{b}}^{*}\Omega_1}{4\Delta_1\Delta_2\delta}\hat{J}_{-}^2
\end{eqnarray}
as shown in Ref.~\cite{anders1}. Tuning the strength of the laser couplings such that $\abs{\Omega_1g_{\text{b}}^{*}}/\Delta_1=\abs{\Omega_2g_{\text{a}}^{*}}/\Delta_2=\abs{\Omega g^{*}}/\Delta$, $\hat{H}_{\text{eff}}$ reduces to the one-axis Hamiltonian $\hat{H}_{\text{1-axis}}=\alpha\hat{J}^2_\theta$ with $\alpha=\abs{\Omega}^2\abs{g}^2/\Delta^2\delta$ and $e^{-2i\theta}=\frac{\Omega_1^{*}g_{\text{b}}g_{\text{a}}^{*}\Omega_2}{\abs{\Omega_1^{*}g_{\text{b}}g_{\text{a}}^{*}\Omega_2}}$~\cite{anders1}. By adding lasers 3 and 4, we basically add the same effective terms to the Hamiltonian as with laser 1 and 2, except they are now proportional to $\Omega_3$, $\Omega_4$, and $-1/\delta$ instead of $\Omega_1, \Omega_2$, and $1/\delta$ (see \figref{fig:figure1}). Matching the strengths of the lasers results in destructive interference of the $\hat{J}_{-}\hat{J}_+$  and $\hat{J}_{+}\hat{J}_-$ terms. In addition, a relative phase of $\pi$ between laser 1 and 3 while laser 2 and 4 are in phase with each other ensures constructive interference of the $\hat{J}_{+}^2$ and $\hat{J}_{-}^2$ terms resulting in an effective two-axis Hamiltonian of the form in Eq.~\eqref{eq:hamil2}. 

We now proceed by deriving the effective Hamiltonian describing the system. Motivated by the above considerations, we assume that we are in the far detuned limit where $\Delta\gg\Omega,\delta, g$. Consequently, we can adiabatically eliminate the excited states of the atoms using the effective operator formalism introduced in Ref.~\cite{florentin}. We neglect fast oscillating terms ($\sim e^{2i\omega_{b} t}$) in the Hamiltonian and assume $1/\Delta_{1}\approx1/(\Delta_{1}+2\delta)$ and $1/\Delta_{2}\approx1/(\Delta_{2}+2\delta)$ since we are considering the limit $\Delta\gg\delta$. After some algebra, we end up with an effective Hamiltonian
\begin{eqnarray} \label{eq:hamileff1}
\hat{H}_{\text{eff1}}\!\!&=&\!-\!\!\!\sum_{k=1}^{N} \nonumber \\ 
&&\!\!\!\left(\frac{(\abs{\Omega_{1}}^{2}\!+\!\abs{\Omega_{3}}^{2})\Delta_{1}}{4\Delta_{1}^{2}+\Gamma^{2}}+\Re\left\{\frac{\Omega^{*}_1\Omega_3e^{2i\delta t}}{4\Delta_{1}-2i\Gamma}\right\}\right)\ket{a}_{k}\bra{a}\! \nonumber \\
&&\!\!\!+\left(\!\frac{(\abs{\Omega_{2}}^{2}\!+\!\abs{\Omega_{4}}^{2})\Delta_{2}}{4\Delta_{2}^{2}+\Gamma^{2}}+\Re\left\{\frac{\Omega^{*}_2\Omega_4e^{2i\delta t}}{4\Delta_{2}-2i\Gamma}\right\}\right)\ket{b}_{k}\bra{b}\nonumber \\
&&\!\!\!+\frac{4\abs{g_{\text{a}}}^{2}\Delta_{2}}{4\Delta_{2}^{2}+\Gamma^{2}}\hat{c}^{\dagger}\hat{c}\ket{a}_{k}\bra{a}+\frac{4\abs{g_{\text{b}}}^{2}\Delta_{1}}{4\Delta_{1}^{2}+\Gamma^{2}}\hat{c}^{\dagger}\hat{c}\ket{b}_{k}\bra{b} \nonumber \\
&&\!\!\!+2\Bigg[\frac{g_{\text{b}}^{*}\Delta_{1}}{\Delta_{1}^{2}+\Gamma^{2}}\left(\Omega_{1}e^{-i\delta t}+\Omega_{3}e^{i\delta t}\right)\hat{c}^{\dagger}\ket{b}_{k}\bra{a} \nonumber \\
&&\!\!\!+\!\frac{g_{\text{a}}^{*}\Delta_{2}}{\Delta_{2}^{2}+\Gamma^{2}}\left(\Omega_{2}e^{-i\delta t}+\Omega_{4}e^{i\delta t}\right)\hat{c}^{\dagger}\ket{a}_{k}\bra{b}\!+\!\text{H.c}\Bigg]\!. \qquad
\end{eqnarray}
The effective Lindblad operators describing the atomic decay are
\begin{eqnarray} \label{eq:lxeff1}
\hat{L}_{x,\text{eff1}}^{(k)}&=&\sqrt{\gamma_{x}}\Bigg[\Bigg(\frac{\Omega_{1}\!+\!\Omega_{3}e^{2i\delta t}}{2\Delta_{1}\!-\!i\Gamma}e^{i \Delta_{1} t}\!+\!\frac{2g_{\text{a}}e^{i(\Delta_{2}+\delta)t}}{2\Delta_{2}\!-\!i\Gamma}\Bigg)\ket{x}_{k}\bra{a}\nonumber \\
&&\!+\!\Bigg(\frac{\Omega_{2}\!+\!\Omega_{4}e^{2i\delta t}}{2\Delta_{2}\!-\!i\Gamma}e^{i \Delta_{2} t}\!+\!\frac{2g_{\text{b}}e^{i(\Delta_{1}\!+\!\delta)t}}{2\Delta_{1}\!-\!i\Gamma}\Bigg)\ket{x}_{k}\bra{b}\Bigg]. 
\end{eqnarray}
The first four terms in Eq.~\eqref{eq:hamileff1} are the AC Stark shifts from the laser fields while the next two terms are the cavity induced shifts of the ground states. The terms $\propto e^{2i\delta t}$ in the AC Stark shifts are fast oscillating for large $\delta$ and can therefore be neglected in this limit. Furthermore, the constant terms can be compensated by properly adjusting the frequency of the laser fields. We will therefore neglect the AC Stark shifts in what follows\footnote{For the one-axis scheme, it is found in Ref.~\cite{anders1} and below, that it may be desirable to operate with $\delta=0$. In this case the AC-Stark shifts can be completely compensated by adjusting the frequency of the laser fields.}. In addition, we also neglect the cavity induced shifts since under the right conditions, these give negligible phase shifts to the ground states as we will discuss later. The remaining terms in Eq.~\eqref{eq:hamileff1} describe Raman transitions from $\ket{a}\to\ket{b}$ ($\ket{b}\to\ket{a}$) through laser 1 or 3 (2 or 4) and the cavity field.

Assuming that the cavity field is weakly populated, we now proceed by adiabatic eliminating the cavity field (see App.~\ref{app:A}). For laser fields tuned such that
\begin{equation} \label{eq:laserfields}
\frac{2\Omega_{1}g_{\text{b}}^{*}\Delta_{1}}{4\Delta_{1}^{2}+\Gamma^{2}}=\frac{2i\Omega_{2}g_{\text{a}}^{*}\Delta_{2}}{4\Delta_{2}^{2}+\Gamma^{2}}=\chi
\end{equation}
and $\Omega_{1}=-\Omega_{3}$, $\Omega_{2}=\Omega_{4}$, we find an effective two-axis twisting Hamiltonian, 
\begin{equation} \label{eq:twoeffect}
\hat{H}_{\text{eff2}}\approx\frac{8i\abs{\chi}^{2}\delta}{4\delta^{2}+\tilde{\kappa}^{2}}\left[\hat{J}_{+}^{2}-\hat{J}_{-}^{2}\right],
\end{equation}
where 
\begin{equation} \label{eq:modrate}
\tilde{\kappa}=\kappa+\Gamma\left(\frac{4\avg{\hat{N}_{a}}\abs{g_{\text{a}}}^{2}}{4\Delta_{2}^{2}+\Gamma^{2}}+\frac{4\avg{\hat{N}_{b}}\abs{g_{\text{b}}}^{2}}{4\Delta_{1}^{2}+\Gamma^{2}}\right)
\end{equation}
is the modified decay rate of the cavity due to the atom-cavity coupling. Here $\hat{N}_{a}=\sum_{k}\ket{a}_{k}\bra{a}$ and $\hat{N}_{b}=\sum_{k}\ket{b}_{k}\bra{b}$ are the atomic number operators. These have been replaced with their average values in deriving the effective dynamics assuming that we can neglect fluctuations around the mean for the calculation of $\tilde{\kappa}$. Note that the effective Hamiltonian in Eq.~\eqref{eq:twoeffect} corresponds to setting $\alpha=16\abs{\chi}^{2}\delta/(4\delta^{2}+\tilde{\kappa}^{2})$ and $\theta=-\pi/4$ in Eq.~\eqref{eq:hamil2}\footnote{By choosing the relative phase between $\Omega_{1}g_{\text{b}}^{*}$ and $\Omega_{2}g_{\text{a}}^{*}$ differently, any generic two-axis Hamiltonian $\propto\left(\hat{J}_{\theta}^2-\hat{J}_{\theta+\frac{\pi}{2}}^2\right)$ can be realized}.
The effective Lindblad operators are
\begin{eqnarray} \label{eq:lindblad}
\hat{L}_{x,\text{eff2}}^{(k)}&=&\sqrt{\gamma_{x}}\Bigg[\frac{\Omega_{1}e^{i\Delta_{1}t}}{2\Delta_{1}-i\Gamma}\left(1-e^{2i\delta t}\right)\ket{x}_{k}\bra{a}\nonumber \\
&&+\frac{\Omega_{2}e^{i\Delta_{2}t}}{2\Delta_{2}-i\Gamma}\left(1+e^{2i\delta t}\right)\ket{x}_{k}\bra{b} \nonumber \\
&&-\left(\frac{2ig_{\text{a}}\chi e^{i\Delta_{2}t}}{2\Delta_{2}-i\Gamma}\ket{x}_{k}\bra{a}+\frac{2g_{\text{b}}\chi e^{i\Delta_{1}t}}{2\Delta_{1}-i\Gamma}\ket{x}_{k}\bra{b}\right)\nonumber \\
&&\times\left(\frac{i\hat{J}_{+}-\hat{J}_{-}}{\delta+i\tilde{\kappa}/2}-\frac{i\hat{J}_{+}+\hat{J}_{-}}{\delta-i\tilde{\kappa}/2}e^{2i\delta t}\right)\Bigg] \label{eq:leff1} \\
\hat{L}_{c,\text{eff2}}&=&-\sqrt{\kappa}\chi e^{-i\delta t}\left[\frac{i\hat{J}_{+}-\hat{J}_{-}}{\delta+i\tilde{\kappa}/2}-\frac{i\hat{J}_{+}+\hat{J}_{-}}{\delta-i\tilde{\kappa}/2}e^{2i\delta t}\right]. \label{eq:leff2}
\end{eqnarray}
We now proceed by deriving the evolution of the collective spin state predicted by the effective operators. 
\subsection{Equations of motion}

The equation of motion (EOM) for the mean of an atomic operator $\avg{\hat{O}}$ can be found from the Heisenberg-Langevin equation
\begin{eqnarray}
\frac{\text{d}}{\text{d}t}\avg{\hat{O}}&=&i\left\langle\left[\hat{H}_{\text{eff2}},\hat{O}\right]\right\rangle+\sum_{x}\sum_{k}\left\langle\left(\hat{L}_{x,\text{eff2}}^{(k)}\right)^{\dagger}\hat{O}\hat{L}_{x,\text{eff2}}^{(k)}\right\rangle\nonumber \\
&&-\frac{1}{2}\left\langle\hat{O}\left(\hat{L}_{x,\text{eff2}}^{(k)}\right)^{\dagger}\hat{L}_{x,\text{eff2}}^{(k)}\right\rangle\nonumber \\
&&-\frac{1}{2}\left\langle\left(\hat{L}_{x,\text{eff2}}^{(k)}\right)^{\dagger}\hat{L}_{x,\text{eff2}}^{(k)}\hat{O}\right\rangle.
\end{eqnarray}
To obtain a closed set of EOMs, we linearize the noise of the atomic operators in the limit of $N\gg1$ similar to what was done in Ref.~\cite{anders1}. The linearization of the noise can be described as making the transformation
\begin{eqnarray}
\hat{J}_{+}&\to&\avg{\hat{J}_{+}}+\lambda\delta\hat{J}_{+}, \qquad \quad \hat{N}_{a}\to\avg{N_{a}} \nonumber \\
\hat{J}_{-}&\to&\avg{\hat{J}_{-}}+\lambda\delta\hat{J}_{-}, \qquad \quad \hat{N}_{b}\to\avg{N_{b}}\nonumber \\
\hat{J}_{z}&\to&\avg{\hat{J}_{z}},
\end{eqnarray}
in the EOMs and only keeping terms to second order in $\lambda$. Here $\delta\hat{O}=\hat{O}-\avg{\hat{O}}$ describe the fluctuations around the mean. The result of this is a closed set of EOMs that can be solved numerically (see Appendix~\ref{app:B}). 

\begin{figure} [h]
\centering
\includegraphics[width=0.47\textwidth]{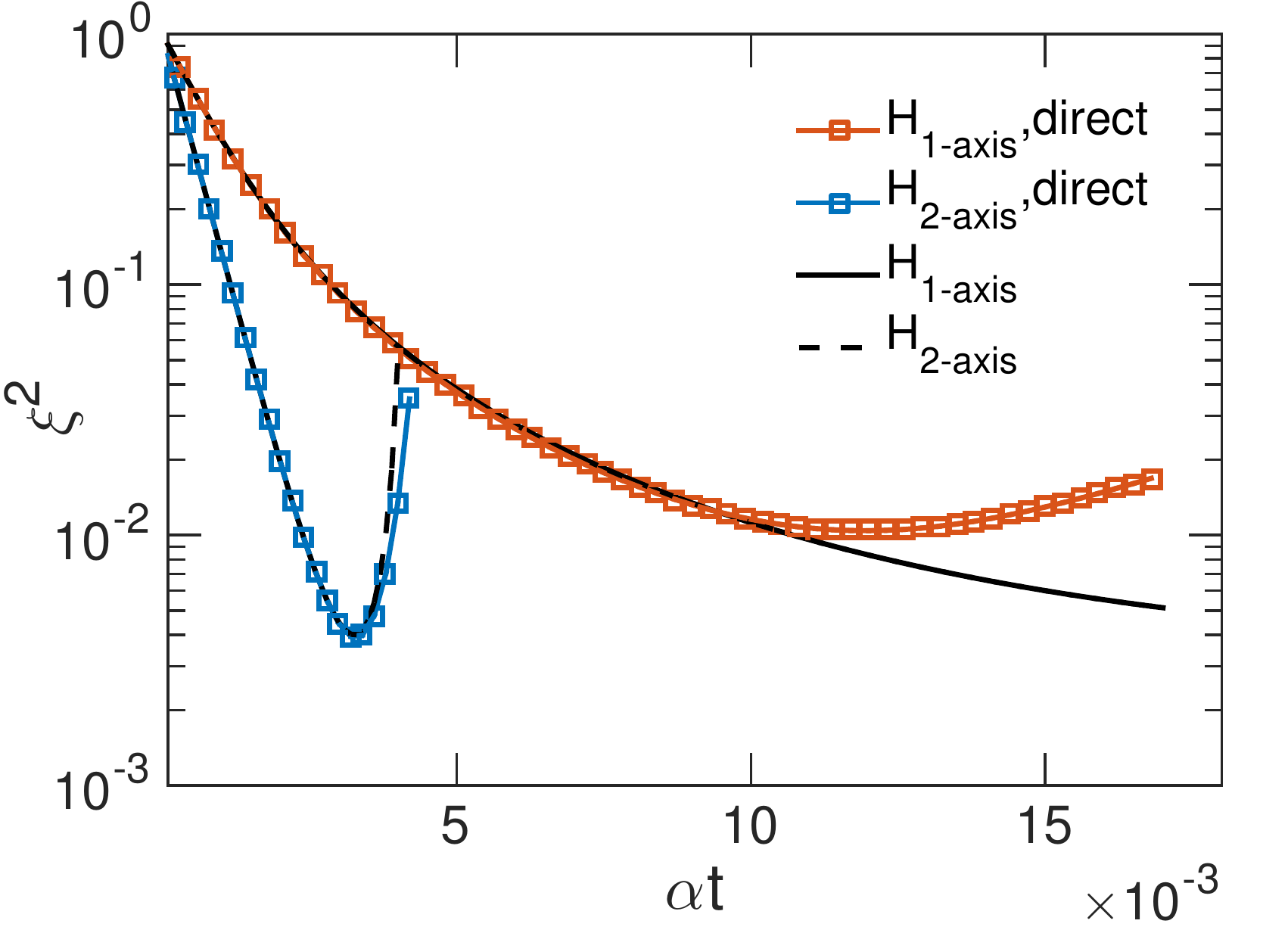}
\caption{Squeezing parameter, $\xi^2$ as a function of time for a two-axis Hamiltonian of the form $\hat{H}_{\text{2-axis}}=\alpha\left(\hat{J}^2_{-\pi/4}-\hat{J}^2_{\pi/4}\right)$ and a one-axis Hamiltonian of the form $\hat{H}_{\text{1-axis}}=\alpha\hat{J}_x^2$. The evolution has been calculated both by direct numerical solution of the Schr\"{o}dinger equation (direct) and by linearization of the noise. For $\hat{H}_{\text{2-axis}}$, the linearization breaks down near the minimum of $\xi^2$ where $\avg{J_z}$ changes significantly from its initial value although it is fairly well described by the linearization. For $\hat{H}_{\text{1-axis}}$ on the other hand, the breakdown happens due to distortions of the squeezing ellipse, which is not contained in the linearized description~\cite{ueda,strobel}. The calculations assume $N=1000$ atoms initially in state $\ket{a}$. }
\label{fig:figureX1}
\end{figure}

In the absence of decoherence, it is also possible to numerically solve the Schr\"{o}dinger equation for a given initial state without performing any linearization of the noise. In order to investigate the accuracy of the linearization performed above, we have therefore evaluated the evolution dictated by a two-axis twisting Hamiltonian of the form $\hat{H}_{\text{2-axis}}=\alpha\left(\hat{J}^2_{-\pi/4}-\hat{J}^2_{\pi/4}\right)$ both by directly solving the Schr\"{o}dinger equation numerically and by performing the linearization of the noise. The squeezing parameter, $\xi^2$ calculated from both methods are shown in \figref{fig:figureX1}.  
We have assumed $N=1000$ atoms and that all atoms start out in state $\ket{a}$. Near the minimum of $\xi^2$, $\hat{J}_z$ begins to decrease rapidly and as a result, our linearization begins to break down. For smaller times, the linearization however captures the dynamics quite accurately. Since dissipation will result in a different minimum of $\xi^2$ at an earlier time than for the bare Hamiltonian evolution, we expect our linearization to be valid also when the squeezing is limited by dissipation. 

For comparison, we have also plotted the squeezing parameter for a one-axis twisting Hamiltonian of the form in Eq.~\eqref{eq:hamil1} using both the direct method and the linearization of the noise. As with the two-axis Hamiltonian, the linearization breaks down near the minimum of $\xi^2$ though the effect is more severe. We believe the reason for this is that the atomic spin is not only squeezed but also twisted into a non-gaussian state for a high amount of squeezing~\cite{davis1,ueda,strobel}. This is not captured by the linearization. Hence the results of the linearization of the one axis Hamiltonian  should not be trusted when the squeezing  is close to the minimum obtained from the Hamiltonian. \figref{fig:figureX1} also shows how the two-axis Hamiltonian results in higher squeezing and squeezes faster than the one-axis Hamiltonian in the absence of decoherence. 

\section{Squeezing analysis}
In order to include the effect of decoherence, we numerically solve the EOMs from the effective operators. First, however, we make some analytical estimates of what to expect in order to better understand the numerical results.  We approximately solve the equations of motion for the evolution of the squeezing parameter $\xi^2$ under $\hat{H}_{\text{eff2}}$, starting from a coherent spin state polarized along $z$.  We assume that dissipation is sufficiently weak that $J = N/2$ is preserved, and we consider the planar limit where $J_z \approx N/2$ throughout the squeezing. From Eqs. \eqref{eq:laserfields} and \eqref{eq:twoeffect}, we find the effective interaction strength $\alpha\approx4\abs{\chi}^2/\delta\approx\Omega^2 g^2/(\Delta^2\delta)$, where we have defined a generic laser coupling $\Omega$, cavity coupling $g$ and detuning $\Delta$ to characterize the system.  We assume the limit of large detuning $\Delta$ where the two first terms of the Lindblad operators $\hat{L}_{x,\text{eff2}}^{(k)}$ describing spontaneous emission in Eq.~\eqref{eq:lindblad} are dominant (see discussion below).

From the Heisenberg-Langevin equation, we find that
\begin{eqnarray}\label{eq:JxSq}
\frac{d\avg{J_x^2}}{dt}&\approx&
-2N\alpha\avg{J_x^2}\nonumber\\
&&+\frac{N^2\kappa\Omega^2g^2}{4\Delta^2\delta^2}+\frac{\Gamma\Omega^2}{8\Delta^2}N,
\end{eqnarray}
where we have assumed that $N\alpha\gg\Gamma\Omega^2/\Delta^2$.  
The resulting evolution of the squeezing parameter is 
\begin{equation}\label{eq:eomxi}
\frac{d\xi^2}{dt}\approx \alpha\left(-2 N \xi^2 +N\frac{\kappa}{\delta} + \frac{1}{2C}\frac{\delta}{\kappa}\right),
\end{equation}
where $C = g^2/(\kappa \Gamma)$ is the single-atom cooperativity.  The first term in Eq.~\eqref{eq:eomxi} is the unitary evolution from the two-axis Hamiltonian while the second and third terms describe noise added by cavity decay and spontaneous emission from the atoms, respectively.  

In the limit of large single-atom cooperativity $C$, where spontaneous emission becomes negligible, dissipation via the cavity can be suppressed by operating at large detuning $\delta$ from cavity resonance.  At finite cooperativity, however, the squeezing is optimized at a detuning $\delta\sim\delta_s=\sqrt{NC}\kappa$ that minimizes the combined effect of the two forms of dissipation.  Squeezing at all requires $d\xi^2 / dt < 0$ at $\xi=1$, corresponding to a large collective cooperativity $\sqrt{NC} >1$.  The squeezing parameter then initially decays until reaching a minimum value of $\xi^2 \sim 1/\sqrt{NC}$ after a time $t_s \sim \ln(\sqrt{NC})/(\alpha N)$, where the rate of squeezing can no longer compete with the rate of adding noise. 

The scaling of the squeezing parameter obtained above is the same as the scaling for the one axis scheme derived in Ref.~\cite{anders1}. One might have expected a more favorable scaling of $\xi^2$ for the two-axis Hamiltonian than for the one-axis Hamiltonian since, in the absence of noise, the former gives an exponential decrease of $\xi^2$ while the latter only leads to a decrease of the form $1/(\alpha N t)$. However, for the coupling configuration leading to the one-axis Hamiltonian $\hat{H}_{\text{1-axis}}=\alpha\hat{J}^2_x$, the effective Lindblad operator describing the cavity decay is $\propto \hat{J}_x$ (see Appendix~\ref{app:A}). For large $\alpha t$, the squeezed component from the Hamiltonian  is almost entirely described by $\hat{J}_x$ with only a small admixture of $\hat{J}_z$. The cavity decay thus nearly conserves the value of the squeezed component and primarily adds noise to the anti-squeezed component $\sim\hat{J}_y$ (\figref{fig:figure0}a).  Consequently, the one-axis scheme is more stable against cavity decay than one would naively expect. For the two-axis Hamiltonian scheme, cavity decay adds noise to both the squeezed and anti-squeezed components (see Appendix~\ref{app:C} and \figref{fig:figure0}b), which counteracts the faster squeezing such that the scaling of $\xi^2$ becomes the same for the two schemes.

Without the assumptions of $J_z=N/2$ and constant $J=N_a+N_b$, we can numerically solve the EOMs given in Appendix~\ref{app:B} to evaluate $\xi^2$ in the limit $N\gg NC$, where the scheme is limited by dissipation.. From the Lindblad operators in Eq.~\eqref{eq:leff1}, we can estimate the effect of spontaneous decay of the atoms on the collective atomic state. To determine the ideal operating  conditions we note that the effect of the two first terms in Eq.~\eqref{eq:leff1} will not decrease with increasing $\Delta_1, \Delta_2$, since we expect $t\propto\Delta^2/(\Omega^2\Gamma)$. The other terms will, however, be suppressed for large detunings ($\Delta$). In the numerical simulations, we find that these terms have a detrimental effect on the squeezing and the detunings should therefore be chosen large enough for these terms to be negligible compared to the two first terms. We include these terms in our numerical optimizations, but choose $\Delta_1, \Delta_2$ sufficiently large that they are negligible. The result are then almost independent of $\Delta_1, \Delta_2$ and we do not optimize over these parameters. 

We numerically minimize $\xi^2$ for the two-axis scheme while requiring the laser fields to be tuned such that $\Omega_2=\Omega_4$ and $\Omega_1=-\Omega_3$ (this assumption forces the Hamiltonian to remain of the two-axis form and not cross into the one-axis Hamiltonian in the numerical optimization). We minimize in the interaction time $t_s$, the two-photon detuning $\delta$, and the ratio $\Omega_2/\Omega_1$ which can be an imaginary number reflecting a phase difference between the two laser fields. Note that we keep $\Omega_{1,2}/\Delta_{1,2}\lesssim1/50$ to ensure the validity of the adiabatic elimination (see below). The result of the optimization is shown in \figref{fig:figure2}(a), which also shows the optimal squeezing for the one-axis scheme considered in Ref.~\cite{anders1}. The effective operators for the one-axis scheme can be obtained from the effective operators of the two-axis scheme by simply setting $\Omega_3=\Omega_4=0$ (see Appendix~\ref{app:A}). The numerical simulations confirm the $1/\sqrt{NC}$ scaling of $\xi^2$ for both schemes and it is seen that the one-axis scheme reaches slightly higher squeezing than the two-axis scheme. 

The numerically optimized detuning and squeezing time for the two-axis scheme are shown in \figref{fig:figure2}(b) and are in good agreement with the analytical estimates of $\delta\sim\sqrt{NC}\kappa$ and $t\sim\frac{\ln(\sqrt{NC})\Delta^2}{\sqrt{NC}\Omega^2}\frac{1}{\Gamma}$. In contrast to this, the maximum squeezing for the one-axis scheme is found for $\delta=0$ where the effective Hamiltonian is vanishing. This was already noted in Ref.~\cite{anders1} but the origin of this result was not clear at the time. From our analysis, we observe that the optimum corresponds to a dissipative scheme, very similar to Ref.~\cite{torre}. In both schemes, the effective Lindblad operator associated with the cavity decay drives the system into a squeezed state for non-balanced laser couplings ($\Omega_2\neq\Omega_1$). The main difference between Ref.~\cite{torre} and the dissipative scheme considered here is that while we consider a three level system with one excited state and two different detunings ($\Delta_1,\Delta_2$), the scheme in Ref.~\cite{torre} considers a four level system with two excited states and equal detunings. 
\begin{figure} [h]
\centering
\subfloat {\label{fig:figure2a}\includegraphics[width=0.5\textwidth]{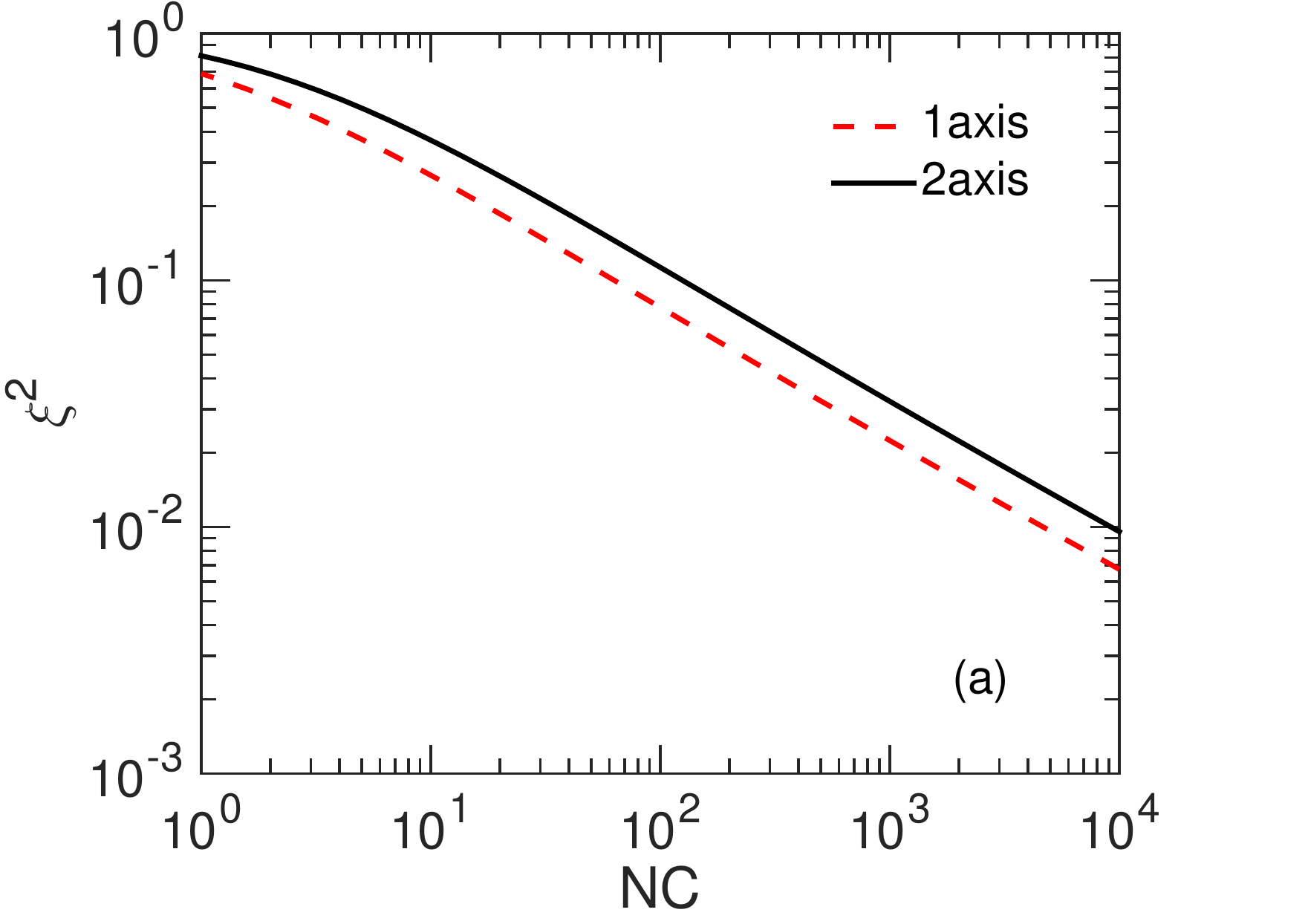}} \\
\subfloat{\label{fig:figure2b}\includegraphics[width=0.5\textwidth]{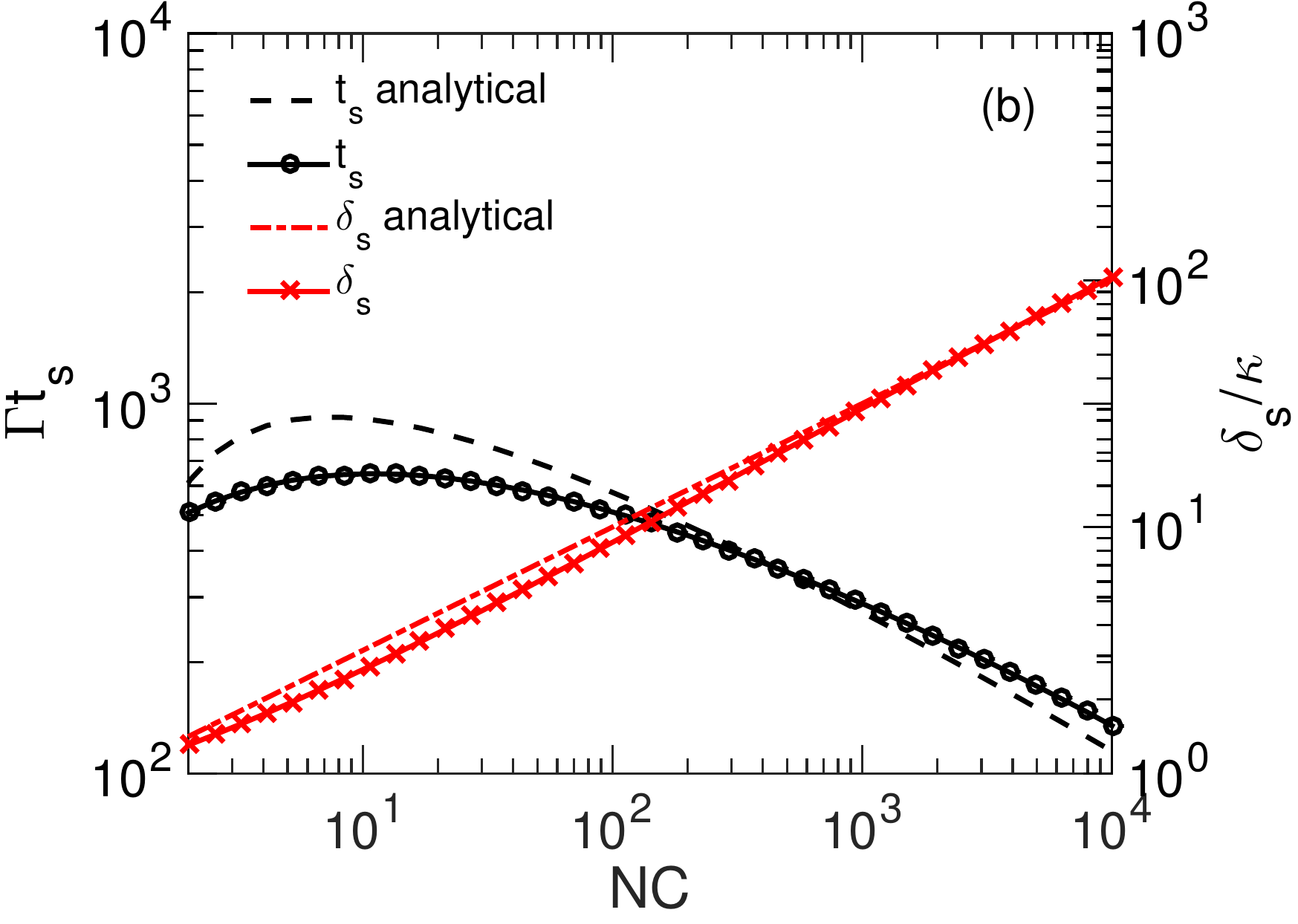}}
\caption{(a) Minimized squeezing parameter $\xi^2$ as a function of the collective cooperativity $NC$ for the two-axis scheme (Two-axis) and the one-axis scheme of Ref.~\cite{anders1} (One-axis). Both schemes exhibit the same $1/\sqrt{NC}$ scaling but the one-axis scheme reaches slightly stronger squeezing than the two-axis scheme. (b) Squeezing time $t_s$ (left axis) and optimal detuning $\delta_s$ (right axis) for the two-axis scheme. The dashed/dot-dashed lines show the analytical estimates of the parameters $\Gamma t_s=\ln(NC)/\sqrt{NC}$ and $\delta_s/\kappa=1/\sqrt{NC}$. For both plots, we have assumed that $\Gamma=\kappa$, $\gamma_a=\gamma_b=\gamma_o$, $g_{\text{a}}=g_{\text{b}}$, $\omega_{b}=10^3\Gamma$, $\Delta_1=20\Gamma NC$, $\Omega_{1,2}/\Delta_{1,2}\sim1/50$, and $N\gg NC$.}
\label{fig:figure2}
\end{figure}
\subsection{Squeezing fidelity}
Even though the degree of squeezing obtainable with the two-axis and one-axis scheme are similar, the squeezing operations are very different. In particular, we found that the one-axis scheme leads to maximum squeezing when operated in a dissipative fashion. To further compare the performance of the two schemes we consider the fidelity of the squeezing operation for both schemes when compared to a perfect squeezing operation on a coherent spin state. We define canonical position and momentum operators~\cite{hammerer1} 
\begin{eqnarray}
\hat{x}&=&\hat{J}_x/\sqrt{\avg{\hat{J}_z}}, \\
\hat{p}&=&\hat{J}_y/\sqrt{\avg{\hat{J}_z}},
\end{eqnarray}
to describe the spin ensemble. We assume that the ensemble is initially prepared in a coherent spin state and that $\avg{\hat{J_z}}\approx N/2\gg1$. In this regime, the canonical operators have the usual canonical commutation relation, $\left[\hat{x},\hat{p}\right]\approx i$ and the spin ensemble is described by a Gaussian state characterized by $\hat{x}$ and $\hat{p}$. The perfect squeezing operation amounts to performing the transformation $\left\{\hat{x},\hat{p}\right\}\to\left\{s\hat{x},\hat{p}/s\right\}$, where $0<s<1$ is squeezing in the $\hat{x}$ quadrature ($s>1$ is squeezing in $\hat{p}$). For a given amount of squeezing, we perform a numerical optimization of the fidelity between the perfectly squeezed Gaussian state and the state produced by either the one-axis or two-axis squeezing scheme. The output states of the squeezing schemes are approximately Gaussian since the initial state is a coherent spin state and we are considering squeezing well above the Heisenberg limit. Since all operations and states are Gaussian they are completely characterized by the first and second moments of $\hat{x}$ and $\hat{p}$. We find these from numerical integration of the EOM as before and calculate the fidelity between the perfectly squeezed state and the output state of the squeezing schemes as described in Ref.~\cite{banchi}. 

We optimize in the interaction time $t_s$, the two-photon detuning $\delta$, and the ratio $\Omega_2/\Omega_1$ assuming that $\Omega_3=-\Omega_1$ and $\Omega_4=\Omega_2$ in the two-axis scheme. We also allow for initial and final rotations of the spin state and optimize in the rotation angles. The unitary evolution from the one-axis and two-axis schemes will, in general, squeeze a linear combination of $\hat{x}$ and $\hat{p}$ depending on the initial state and the amount of squeezing. Initial and final rotations ensure that the output state is squeezed in $\hat{x}$. In the previous squeezing analysis, we simply considered an initial state at the origin ($\avg{\hat{x}}=\avg{\hat{p}}=0$) in our numerical simulations since we were only interested in the maximal amount of squeezing obtainable with the two schemes. However, if the squeezing operation is to be performed as part of a continuous variable quantum information protocol~\cite{rev123} or to enhance the measured signal in a quantum metrology protocol~\cite{davis1,hosten,linnemann2016}, the initial state will in general not be at the origin. We therefore consider a distribution of initial states around the origin to ensure a fair comparison. We consider rotation symmetric distributions of initial states with equal distance to the origin, $r=\sqrt{\avg{\hat{x}}^2+\avg{\hat{p}}^2}$ and calculate the average squeezing fidelity for a given distance. 

\begin{figure}
\centering
\subfloat{\label{fig:figure3a}\includegraphics[width=0.46\textwidth]{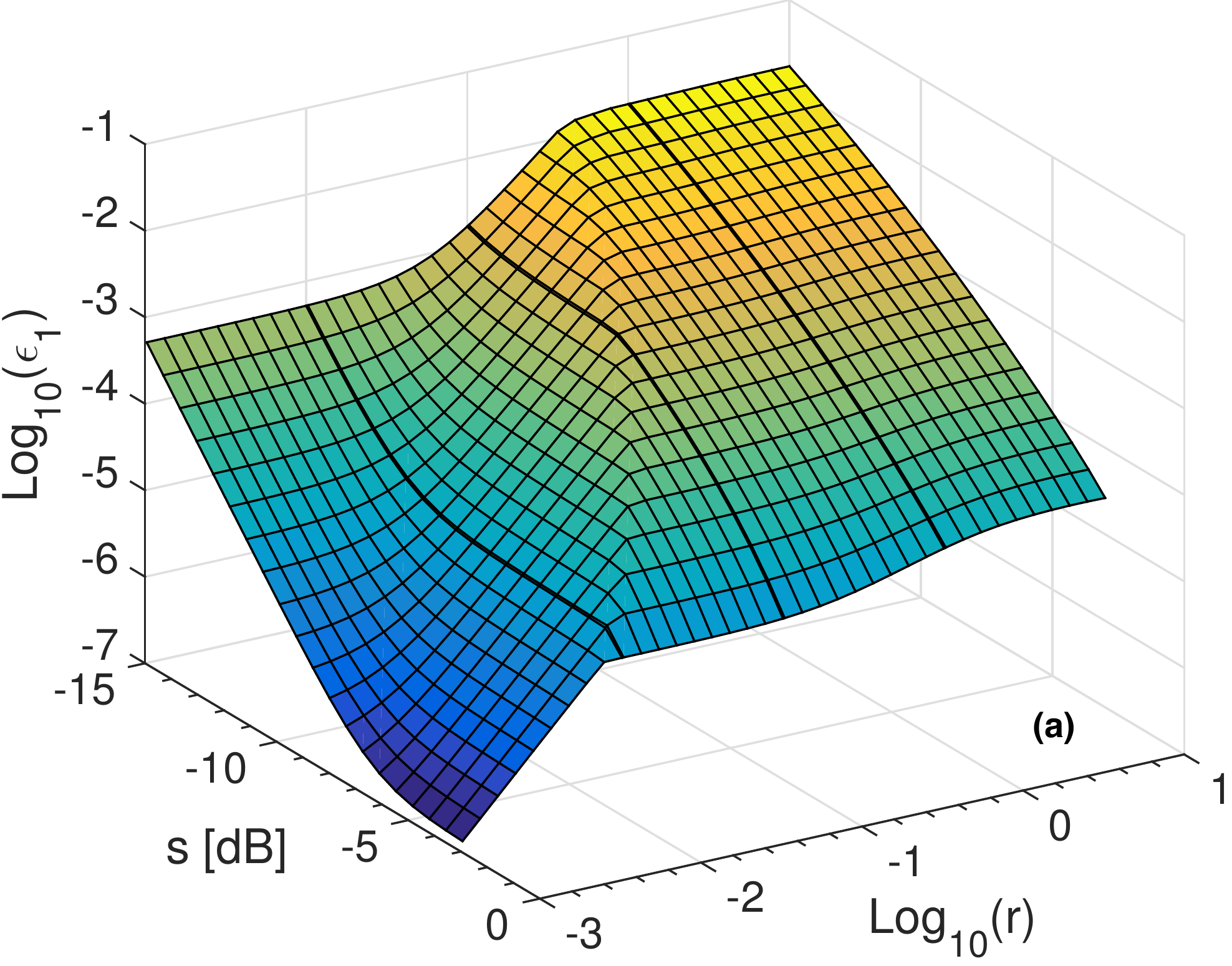}} \\
\subfloat{\label{fig:figure3b}\includegraphics[width=0.46\textwidth]{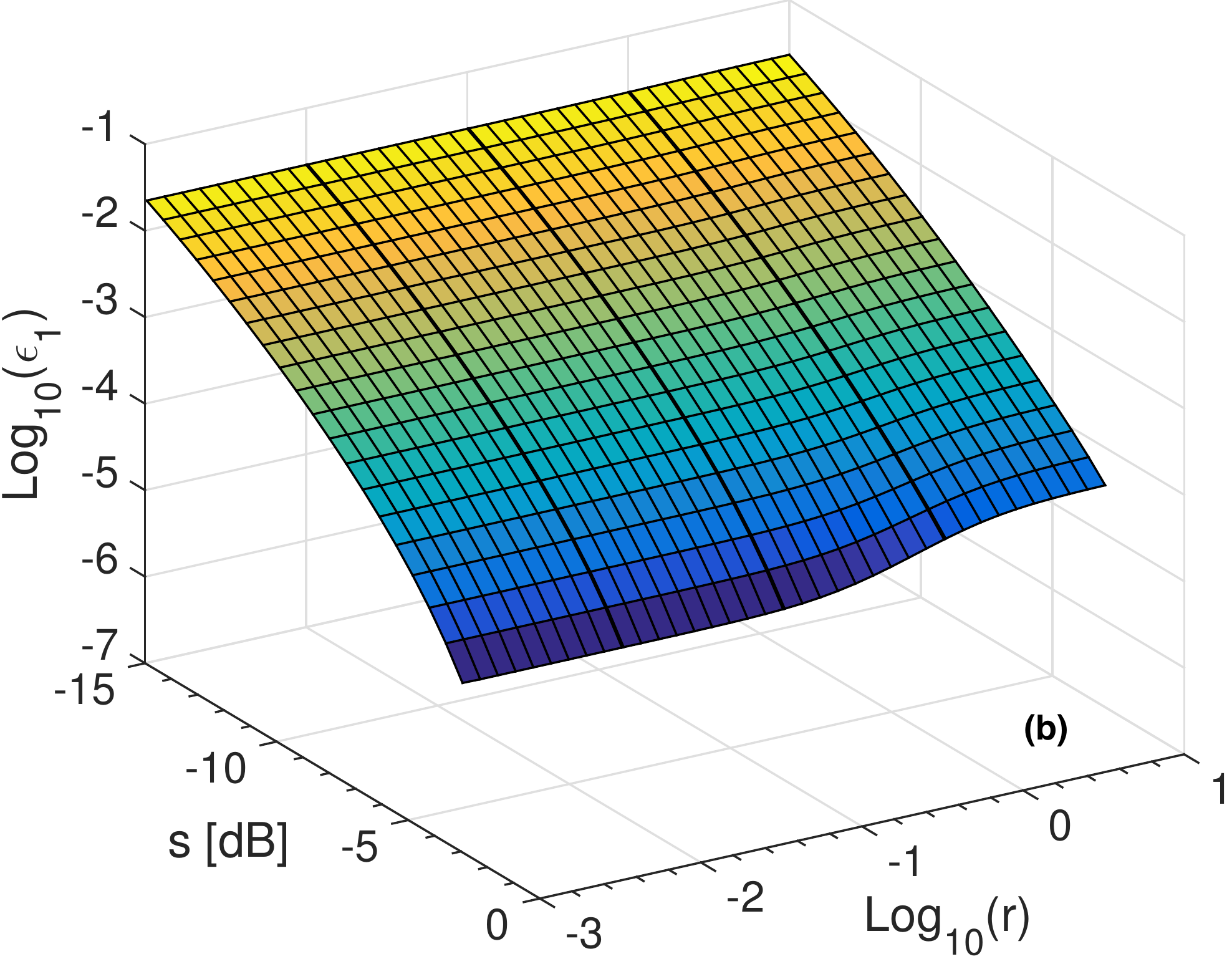}} \\
\subfloat{\label{fig:figure3c}\includegraphics[width=0.46\textwidth]{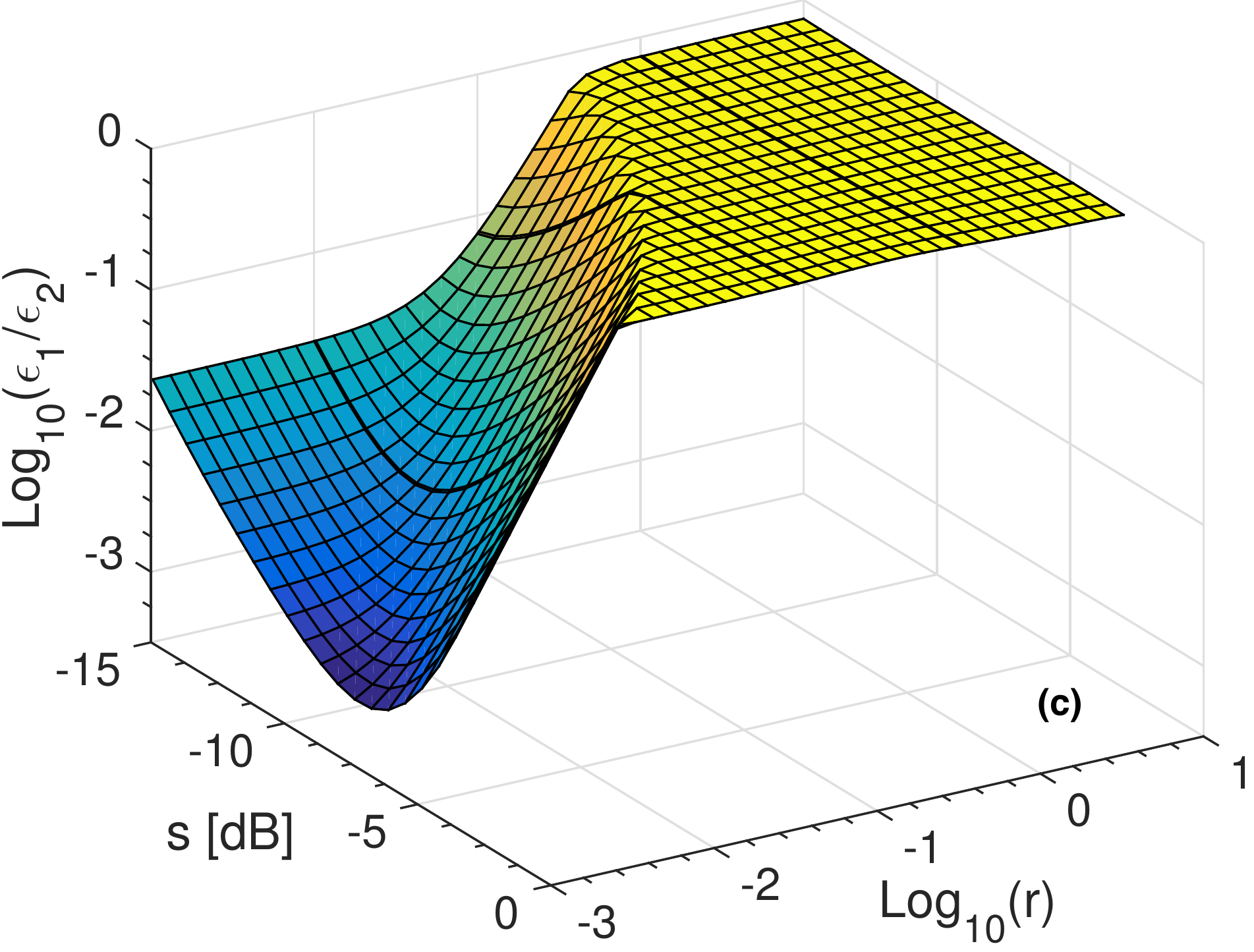}}
\caption{Errors for the one-axis (a) and two-axis (b) squeezing schemes and the ratio of these (c) as a function of the squeezing parameter $s$ and canonical displacement $r$ of the initial state. The errors are defined as $\epsilon=1-F$ where $F$ is the fidelity of the output state with a perfectly squeezed state. The transition from dissipative to unitary operation of the one-axis scheme is seen in (a) and (c) as the transition between the regions with strong and weak $r$-dependence. We have assumed that $\Gamma=\kappa$, $\gamma_a=\gamma_b=\gamma_o$, $g_{\text{a}}=g_{\text{b}}$, $\omega_{b}=10^3\Gamma$, $\Delta_1=20\Gamma NC$, $\Omega_{1,2}/\Delta_{1,2}\sim1/50$ and $N\gg NC=1000$. $\text{Log}_{10}$ refers to the logarithm with base 10.}
\label{fig:figure3}
\end{figure}

In \figref{fig:figure3}(a) and (b) we show the resulting infidelity $\epsilon=1-F$ for the one- and two-axis Hamiltonians. In \figref{fig:figure3}(c), we compare the ratio of the infidelities.  For both approaches the infidelity vanishes for small squeezing $\log(s) \to 0$, where the system is almost unperturbed, and increases as the squeezing is increased. For the two-axis scheme, the error is almost independent of the displacement, whereas the one axis scheme display a cross over between two modes of operation. For small displacements the dissipative approach yields a much better performance, whereas for large displacements unitary operation is desirable resulting in a fidelity almost independent of the size of the displacement.

The reason for this is that the dissipative scheme drives the system towards a squeezed dark state at the origin i.e. it decreases both $\avg{\hat{x}}$ and $\avg{\hat{p}}$ and as a result, the fidelity with the perfectly squeezed state decreases as we move away from the origin. This is not the case for the unitary one-axis scheme, which therefore performs better away from the origin and leads to similar but slightly better fidelities than the two-axis scheme.
The relative performance of the one-axis and two-axis scheme might change if one moves out of the planar limit considered here where $\avg{\hat{J}_z}\approx N/2$. Since the one-axis Hamiltonian distorts the state quite severely for high squeezing ~\cite{davis1}, it might perform worse than the two-axis Hamiltonian in this limit when compared to an ideal squeezing operation. As the initial state moves away from the pole of the Bloch sphere, this effect might also become more severe. It is, however, beyond the scope of this work to consider the non-planar limit.  

\subsection{Validity of effective dynamics}
\begin{figure} [h]
\centering
\includegraphics[width=0.47\textwidth]{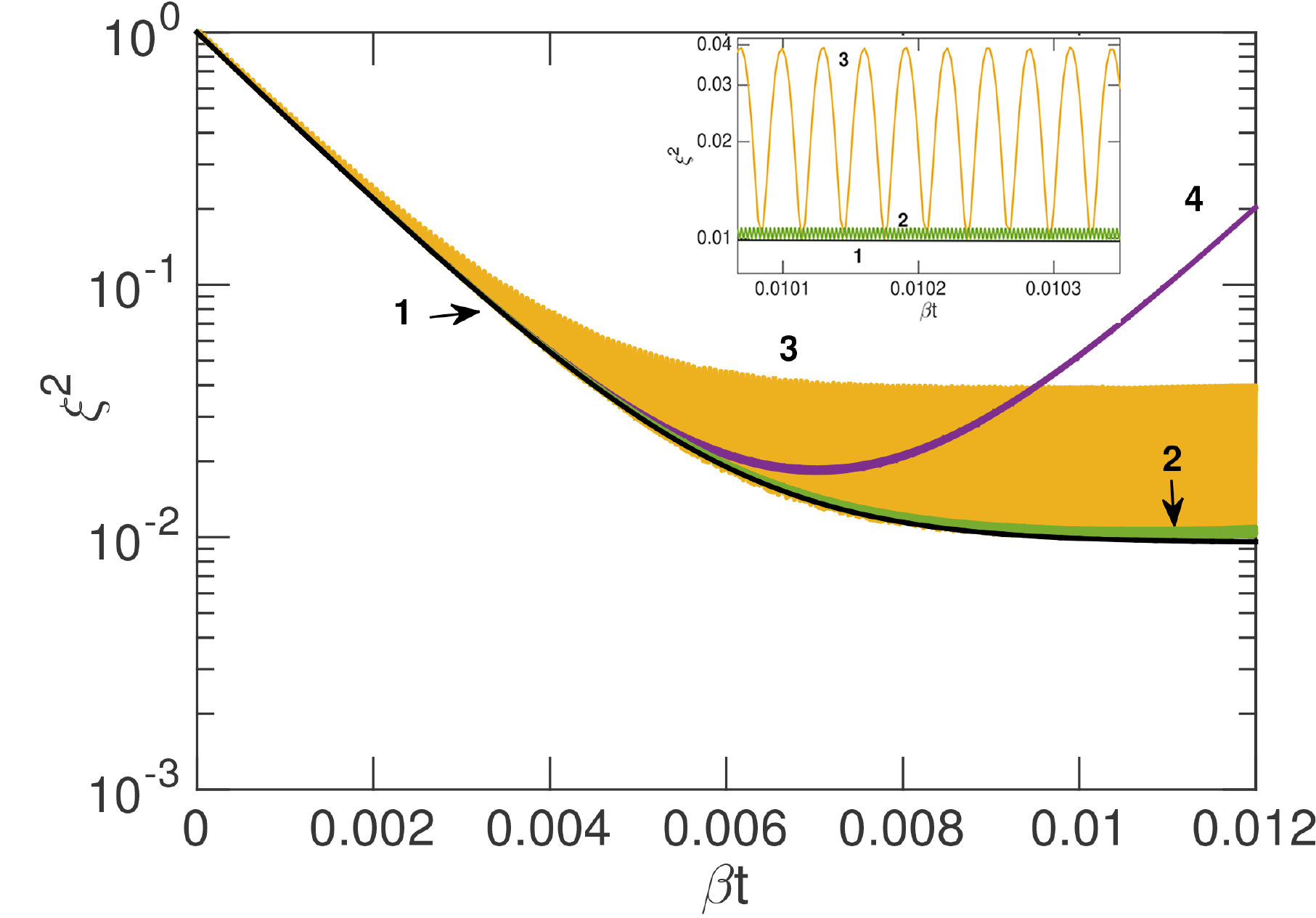}
\caption{Evolution of the squeezing parameter $\xi^2$ as a function of the rescaled time $\beta t$, where $\beta=2\Gamma\Omega_1^2\Delta_1^2/(4\Delta_1^2+\Gamma^2)^2=2\Gamma\Omega_2^2\Delta_2^2/(4\Delta_2^2+\Gamma^2)^2$. Curve 1 (black) is calculated for the model where the cavity has been eliminated for $\Delta_{1,2}/\Omega_{1,2}\sim 50$ and $\Delta_{1}=20NC$ while curves 2 (green), 3 (orange), and 4 (purple) have been calculated for the cavity model with parameters $\Delta_{1,2}/\Omega_{1,2}\sim 50, 16, 50$ and $\Delta_{1}=20NC, 20NC, 2NC$, respectively. Curve 2 satisfies the criterion for adiabatic elimination while curves 3 and 4 violate the condition on $\Omega$ and $\Delta$, respectively. The effect of too strong driving is that $\xi^2$ has large oscillations on a timescale of $1/\delta$ as shown by curve 3. Hence the seemingly colored orange and green areas are not colored, but are curves with very fast oscillations (see inset). Curve 4 shows how the cavity induced shift introduce a deviation from the desired behavior resulting in a weaker squeezing $\xi^2$ when $\Delta$ is to small to neglect cavity shifts. The calculations were performed with the same parameters as in \figref{fig:figure2} for $NC=10^4$ and $\delta=100\kappa\sim\delta_s$ assuming the laser fields to be tuned according to Eq.~\eqref{eq:laserfields}.}
\label{fig:figure4}
\end{figure}

In our analysis so far, we have neglected the cavity shifts in the effective Hamiltonian $\hat{H}_{\text{eff1}}$ in Eq.~\eqref{eq:hamileff1}. Furthermore, we have neglected terms oscillating as $e^{2i\delta t}$ or faster in the EOMs of the two-axis scheme. In order to investigate these assumptions, we have performed numerical simulations using $\hat{H}_{\text{eff1}}$ and $\hat{L}_{x,\text{eff1}}^{(k)}$ without adiabatically eliminating the cavity field and only neglecting terms oscillating faster than $e^{2i\delta t}$. We still neglect the AC Stark shifts $\propto \abs{\Omega}^2$ in $\hat{H}_{\text{eff1}}$ since the constant part of these can be compensated by adjusting the frequencies of the lasers, and it is clear that the fast oscillating terms are negligible for weak driving. We do, however, keep the cavity induced shifts where the requirements for negligible shifts are more subtle. We perform a linearization of the noise as before, but now include the cavity field operator in the transformation.  This allows us to obtain a closed set of EOMs for the same mean values as before, and also for $\avg{\hat{J}_{+}\hat{c}},\avg{\hat{J}_{+}\hat{c}^{\dagger}}, \avg{\hat{c}^{2}}$,$\avg{\hat{c}^{\dagger}\hat{c}}$ and their Hermitian conjugates. These expressions thus replace the EOMs presented in Appendix~\ref{app:B} but for brevity, we do not reproduce them here. From these equations, we can investigate under which conditions the adiabatic elimination of the cavity field is valid.  
We find that a sufficient condition to neglect the cavity shifts is that $\Delta_{1,2}\gg NC \Gamma$ and furthermore, we need $1 \gg 8N\abs{\chi}^2\delta/(4\delta^2+\kappa^2)$ in order for the adiabatic elimination of the cavity to be valid. For $\delta=\delta_s$ the latter criterion translates into $(\Gamma/\kappa)\abs{\Omega_{1,2}}^2/\Delta_{1,2}^2\ll1/\sqrt{NC}$. Thus, we can always ensure the validity of the adiabatic elimination if we keep the dynamics slow enough using sufficiently weak laser fields and large detunings $\Delta_{1,2}$. \figref{fig:figure4} shows how the model where the cavity field has been adiabatically eliminated compares to the one without the adiabatic elimination confirming the above conclusion.    

\section{Conclusion and discussion}

We have shown how an effective two-axis twisting Hamiltonian can be realized with a collection of atoms inside an optical cavity. The resulting dynamics of this Hamiltonian leads to spin squeezing of the atoms and, in the absence of dissipation, reaches the ideal Heisenberg limit for metrology. However, the maximum squeezing obtainable in the presence of dissipation is similar to what was found for the one-axis twisting Hamiltonian in Ref.~\cite{anders1} and scales as $1/\sqrt{NC}$, where $NC$ is the collective cooperativity.   The reason why the two-axis scheme does not squeeze more strongly than the one-axis scheme---despite squeezing more quickly---is that collective decay through the cavity mode adds significantly more noise to the squeezed quadrature in the two-axis scheme.  It is therefore expected that if the collective decay can be suppressed the two-axis scheme would outperform the one-axis scheme.  This motivates schemes without strong collective decay, such as squeezing through the Rydberg blockade~\cite{klaus2}.

We have furthermore compared the fidelities of both the one-axis scheme and the two-axis scheme with respect to a perfect squeezing operation. We found that a dissipative operation of the one-axis scheme performed significantly better than the two-axis scheme for squeezing of a state near the origin in phase space. Away from the origin, however, the unitary versions of both the one-axis and two-axis schemes outperform the dissipative scheme and lead to similar fidelities.  Unitary versions may therefore be desirable for continuous variable quantum information processing~\cite{rev123} or surpassing detection noise in quantum metrology~\cite{davis1,hosten,linnemann2016}.  Approximately unitary dynamics might be realized by quantum erasure schemes \cite{trail2010,leroux2012} or with small ensembles in the ultra-strong coupling regime $C\sim N$.

In our analysis, we have assumed that all atoms are equally coupled to the cavity field. This motivates a realization where the atoms are trapped in an optical lattice as demonstrated in Refs.~\cite{monika1,leroux2012,jongmin,bohnet}. For uneven couplings, if the atoms are subject to a uniform driving $\Omega$ and do not move, the dynamics are expected to resemble the homogeneous case and we thus expect similar results also for inhomogeneous coupling~\cite{hammerer1,hosseini1}.  An extension of this work would be to include fluctuating couplings of the atoms, which e.g. would be the case for systems where a large atomic ensembles is trapped inside a glass cell~\cite{vasilakis}. Such systems can contain many millions of atoms, which could compensate a smaller coupling to the cavity field since the relevant parameter is the collective cooperativity $NC$. Furthermore, by allowing the atoms to transverse the beam sufficiently many times during the interaction, one can obtain a motional averaging such that the interaction is effectively with the symmetric collective mode despite the random positions of the atoms~\cite{borregaard3}.  As a result, large atomic squeezed states could be realized. 

During the preparation of this manuscript, we became aware of related work on realizing a two-axis twisting Hamiltonian in a cavity setup~\cite{zhang123,hu123}. The setup described in Ref.~\cite{zhang123} is very similar to ours and they also find Heisenberg limited squeezing when not limited by dissipation. In contrast to Ref.~\cite{zhang123}, however, we also analyse the performance of the scheme in the presence of strong dissipation and find a similar performance as for the one-axis scheme. The scheme in Ref.~\cite{hu123} is related in the sense that the squeezing operation can be described by an effective two-axis squeezing operation but the mechanism is quite different from what is described here and the squeezing is assumed to happen on an optical transition. Nevertheless, this work obtains the same scaling of the squeezing parameter as $1/\sqrt{NC}$.      

\begin{acknowledgements}
We gratefully acknowledge the support from the Carlsberg Foundation, the European Research Council under the European Union's Seventh Framework Programme (FP/2007-2013) through ERC Grant QIOS (Grant No. 306576) and ERC Grant Agreement no 337603, the Danish Council for Independent Research (Sapere Aude), Qubiz - Quantum Innovation Center, and VILLUM FONDEN via the QMATH Centre of Excellence (Grant No. 10059). G. B. and E.D. acknowledge support from the National Science Foundation.  E.D. acknowledges support from the Hertz Foundation.  M. S.-S. acknowledges support from the Alfred P. Sloan Foundation and AFOSR.
\end{acknowledgements}  

\appendix
\section{Effective operators} \label{app:A}
Here we give the effective operators describing the dynamics after adiabatically eliminating the excited states of the atoms and the cavity field. We have neglected the AC Stark shifts and cavity shifts of the atomic ground states as described in the main text. Furthermore, we have not assumed that the laser fields are tuned according to Eq.~\eqref{eq:laserfields}. The effective Hamiltonian is
\begin{widetext}
\begin{eqnarray}
\hat{H}_{\text{eff2}}&=&-\frac{1}{2}\Bigg(\frac{\abs{g_{\text{b}}}^2\Delta_1^2}{\left(2\Delta_1^2+\Gamma^2/2\right)^2}\left(\frac{\abs{\Omega_3}^2}{\delta-i\tilde{\kappa}/2}-\frac{\abs{\Omega_1}^2}{\delta+i\tilde{\kappa}/2}\right)\hat{J}_{+}\hat{J}_{-}+\frac{\abs{g_{\text{a}}}^2\Delta_2^2}{\left(2\Delta_2^2+\Gamma^2/2\right)^2}\left(\frac{\abs{\Omega_4}^2}{\delta-i\tilde{\kappa}/2}-\frac{\abs{\Omega_2}^2}{\delta+i\tilde{\kappa}/2}\right)\hat{J}_{-}\hat{J}_{+}+\nonumber \\
&&\frac{g_{\text{a}}g^{*}_b\Delta_1\Delta_2}{\left(2\Delta_1^2+\Gamma^2/2\right)\left(2\Delta_2^2+\Gamma^2/2\right)}\left(\frac{\Omega_3\Omega_4^{*}}{\delta-i\tilde{\kappa}/2}-\frac{\Omega_1\Omega_2^{*}}{\delta+i\tilde{\kappa}/2}\right)\hat{J}_{-}\hat{J}_{-}\nonumber \\
&&+\frac{g^{*}_ag_{\text{b}}\Delta_1\Delta_2}{\left(2\Delta_1^2+\Gamma^2/2\right)\left(2\Delta_2^2+\Gamma^2/2\right)}\left(\frac{\Omega^{*}_3\Omega_4}{\delta-i\tilde{\kappa}/2}-\frac{\Omega^{*}_1\Omega_2}{\delta+i\tilde{\kappa}/2}\right)\hat{J}_{+}\hat{J}_{+}\Bigg)+\text{H. c.}
\end{eqnarray}
\end{widetext}
The effective Lindblad operators describing atomic decay are
\begin{widetext}
\begin{eqnarray}
\hat{L}^{(k)}_{x2,\text{eff2}}&=&\sqrt{\gamma_x}\Bigg(\frac{\Omega_1e^{i\Delta_1t}+\Omega_3e^{i\Delta_3t}}{2\Delta_1-i\Gamma}\ket{x}_k\bra{a}+\frac{\Omega_2e^{i\Delta_2t}+\Omega_4e^{i\Delta_4t}}{2\Delta_2-i\Gamma}\ket{x}_k\bra{b}\nonumber \\
&&-\frac{\abs{g_{\text{a}}}^2}{2\Delta_2^2+\Gamma^2/2}\frac{\Delta_2}{\Delta_2-i\Gamma/2}\left(\frac{\Omega_4e^{2i\delta t}}{\delta-i\tilde{\kappa}/2}-\frac{\Omega_2}{\delta+i\tilde{\kappa}/2}\right)e^{i\Delta_2t}\ket{x}_k\bra{a}\hat{J}_{+}\nonumber \\
&&-\frac{\abs{g_{\text{b}}}^2}{2\Delta_1^2+\Gamma^2/2}\frac{\Delta_1}{\Delta_1-i\Gamma/2}\left(\frac{\Omega_3e^{2i\delta t}}{\delta-i\tilde{\kappa}/2}-\frac{\Omega_1}{\delta+i\tilde{\kappa}/2}\right)e^{i\Delta_1t}\ket{x}_k\bra{b}\hat{J}_{-}\nonumber \\
&&-\frac{g_{\text{a}}g_{\text{b}}^{*}}{2\Delta_1^2+\Gamma^2/2}\frac{\Delta_1}{\Delta_2-i\Gamma/2}\left(\frac{\Omega_3e^{2i\delta t}}{\delta-i\tilde{\kappa}/2}-\frac{\Omega_1}{\delta+i\tilde{\kappa}/2}\right)e^{i\Delta_2t}\ket{x}_k\bra{a}\hat{J}_{-}\nonumber \\
&&-\frac{g_{\text{a}}^{*}g_{\text{b}}}{2\Delta_2^2+\Gamma^2/2}\frac{\Delta_2}{\Delta_1-i\Gamma/2}\left(\frac{\Omega_4e^{2i\delta t}}{\delta-i\tilde{\kappa}/2}-\frac{\Omega_2}{\delta+i\tilde{\kappa}/2}\right)e^{i\Delta_1t}\ket{x}_k\bra{b}\hat{J}_{+}\Bigg),
\end{eqnarray}
\end{widetext}
while the effective Lindblad operator describing the cavity decay is
\begin{widetext}
\begin{equation}
\hat{L}_{c,\text{eff2}}=\sqrt{\kappa}\left(\frac{g_{\text{b}}^{*}\Delta_1}{2\Delta_1^2+\Gamma^2/2}\left(\frac{\Omega_1e^{-i\delta t}}{\delta+i\tilde{\kappa}/2}-\frac{\Omega_3e^{i\delta t}}{\delta-i\tilde{\kappa}/2}\right)\hat{J}_{-}+\frac{g_{\text{a}}^{*}\Delta_2}{2\Delta_2^2+\Gamma^2/2}\left(\frac{\Omega_2e^{-i\delta t}}{\delta+i\tilde{\kappa}/2}-\frac{\Omega_4e^{i\delta t}}{\delta-i\tilde{\kappa}/2}\right)\hat{J}_{+}\right)
\end{equation}
\end{widetext}
\section{Equations of motion} \label{app:B}
Here we give the expression for the linearized equations of motion for the model where the cavity field has been eliminated. Note that we have neglected all terms oscillating as $e^{2i\delta t}$ or faster in the two-axis scheme. The validity of this is discussed in the main text. In order to simplify the expressions, we write the effective operators defined in Appendix~\ref{app:A} as
\begin{widetext}
\begin{eqnarray}
\hat{H}_{\text{eff2}}&=&-\frac{1}{2}\left(H_{+-}\hat{J}_{+}\hat{J}_{-}+H_{++}\hat{J}_{+}\hat{J}_{+}+\text{H. c.}\right), \\
\hat{L}^{(k)}_{x,\text{eff2}}&=&\sqrt{\gamma_x}\left(\chi_1\ket{x}_k\bra{a}+\chi_2\ket{x}_k\bra{b}+\chi_3\ket{x}_k\bra{a}\hat{J}_{+}+\chi_4\ket{x}_k\bra{b}\hat{J}_{-}+\chi_5\ket{x}_k\bra{a}\hat{J}_{-}+\chi_6\ket{x}_k\bra{b}\hat{J}_{+}\right), \\
\hat{L}_{c,\text{eff2}}&=&\kappa_1\hat{J}_{-}+\kappa_2\hat{J}_{+}. 
\end{eqnarray}
\end{widetext}
With these definitions, the EOMs are
\begin{widetext}
\begin{eqnarray}
\frac{d}{dt}\avg{\hat{J}^2_{+}}&=&2\avg{\hat{J}_z}\left(\left(iH_{++}^{*}-\kappa_1\kappa_2^*\right)\avg{\hat{J}_{+}\hat{J}_{-}}+\left(iH_{++}^{*}+\kappa_1\kappa_2^*\right)\avg{\hat{J}_{-}\hat{J}_{+}}+\left(2i\Re\{H_{+-}\}+\abs{\kappa_1}^2-\abs{\kappa_2}^2\right)\avg{\hat{J}^2_{+}}\right)\nonumber \\
&&-\Gamma\Bigg(\left(\abs{\chi_1}^2+\abs{\chi_2}^2\right)\avg{\hat{J}^2_+}+\left(\abs{\chi_3}^2+\abs{\chi_6}^2\right)\left(3\avg{\hat{J}_+^2}|\avg{\hat{J}_+}|^2+3\avg{\hat{J}_-\hat{J}_+}\avg{\hat{J}_+}^2-5|\avg{\hat{J}_+}|^2\avg{\hat{J}_+}^2\right)\nonumber \\
&&+\left(\abs{\chi_4}^2+\abs{\chi_5}^2\right)\left(3|\avg{\hat{J}_+}|^2\avg{\hat{J}_+^2}+3\avg{\hat{J}_+}^2\avg{\hat{J}_+\hat{J}_-}-5\avg{\hat{J}_+}^2|\avg{\hat{J}_+}|^2\right)\nonumber \\
&&+2\avg{\hat{J}_z}\avg{\hat{J}^2_+}\left(\avg{\hat{N}_a}\left(\abs{\chi_3}^2-\abs{\chi_5}^2\right)+\avg{\hat{N}_b}\left(\abs{\chi_6}^2-\abs{\chi_4}^2\right)\right)+\chi_1\chi_6^*\left(\left(\avg{\hat{N}_a}+\avg{\hat{N}_b}\right)\avg{\hat{J}_-\hat{J}_+}+2\avg{\hat{J}_z}\avg{\hat{J}_+\hat{J}_-}\right)\nonumber \\
&&+\left(2\left(\chi_2\chi_3^*-\chi_1^*\chi_4\right)\avg{\hat{J}_z}+\left(\chi_1\chi_4^*+\chi_2^*\chi_3\right)\left(\avg{\hat{N}_b}+\avg{\hat{N}_a}\right)\right)\avg{\hat{J}^2_+}\nonumber \\
&&+\chi_2^*\chi_5\left(\left(\avg{\hat{N}_a}+\avg{\hat{N}_b}\right)\avg{\hat{J}_+\hat{J}_-}-2\avg{\hat{J}_z}\avg{\hat{J}_-\hat{J}_+}\right)+\left(\chi_3\chi_5^*+\chi_4^*\chi_6\right)\left(6\avg{\hat{J}_+}^2\avg{\hat{J}_+^2}-5\avg{\hat{J}_+}^4\right)\nonumber \\
&&+\left(\chi_3^*\chi_5+\chi_4\chi_6^*\right)\left(\avg{\hat{J}_+}^2\avg{\hat{J}_-^2}+\avg{\hat{J}_-}^2\avg{\hat{J}_+^2}+2|\avg{\hat{J}_+}|^2\left(\avg{\hat{J}_+\hat{J}_-}+\avg{\hat{J}_-\hat{J}_+}\right)-5|\avg{\hat{J}_+}|^4\right)\Bigg)\\
\frac{d}{dt}\avg{\hat{J}_+\hat{J}_-}&=&4\avg{\hat{J}_z}\Im\left\{H_{++}\avg{\hat{J}^2_+}\right\}+2\left(\abs{\kappa_1}^2\avg{\hat{J}_+\hat{J}_-}-\abs{\kappa_2}^2\avg{\hat{J}_-\hat{J}_+}\right)\avg{\hat{J}_z}-\Gamma\Bigg(\left(\abs{\chi_1}^2+\abs{\chi_2}^2\right)\avg{\hat{J}_+\hat{J}_-}\nonumber \\
&&-\abs{\chi_2}^2\hat{N}_a+\left(\abs{\chi_5}^2+\abs{\chi_4}^2\right)\left(\avg{\hat{J}_-}^2\avg{\hat{J}_+^2}+\avg{\hat{J}_+}^2\avg{\hat{J}_-^2}+4|\avg{\hat{J}_+}|^2\avg{\hat{J}_+\hat{J}_-}-5|\avg{\hat{J}_+}|^4\right)\nonumber \\
&&-2\left(\abs{\chi_5}^2\avg{\hat{N}_a}+\abs{\chi_4}^2\avg{\hat{N}_b}\right)\avg{\hat{J}_z}\avg{\hat{J}_+\hat{J}_-}-\abs{\chi_4}^2\avg{\hat{N}_a}\avg{\hat{J}_+\hat{J}_-}-\abs{\chi_6}^2\avg{\hat{N}_a}\avg{\hat{J}_-\hat{J}_+} \nonumber \\ 
&&+\left(\abs{\chi_6}^2+\abs{\chi_3}^2\right)\left(\avg{\hat{J}_-}^2\avg{\hat{J}_+^2}+\avg{\hat{J}_+}^2\avg{\hat{J}_-^2}+|\avg{\hat{J}_+}|^2\left(\avg{\hat{J}_+\hat{J}_-}+3\avg{\hat{J}_-\hat{J}_+}\right)-5|\avg{\hat{J}_+}|^4\right) \nonumber \\
&&+2\left(\abs{\chi_3}^2\avg{\hat{N}_a}+\abs{\chi_6}^2\avg{\hat{N}_b}\right)\avg{\hat{J}_z}\avg{\hat{J}_-\hat{J}_+}+2\Re\left\{\chi_1^*\chi_6\avg{\hat{J}_+^2}\right\}\left(\avg{N_a}-1\right)\nonumber \\
&&+2\Re\left\{\chi_1^*\chi_4\right\}\left(\avg{\hat{N}_b}-1\right)\avg{\hat{J}_+\hat{J}_-}+2\Re\left\{\chi_2^*\chi_3\right\}\left(\avg{\hat{N}_a}-1\right)\avg{\hat{J}_-\hat{J}_+}+2\Re\left\{\chi_2\chi_5^*\avg{\hat{J}_-^2}\right\}\left(\avg{\hat{N}_b}-1\right)\nonumber \\
&&-2\Re\left\{\left(\chi_4^*\chi_6+\chi_3\chi_5^*\right)\left(\avg{\hat{J}_+}^2\left(\avg{\hat{J}_-\hat{J}_+}+2\avg{\hat{J}_+\hat{J}_-}\right)+3|\avg{\hat{J}_+}|^2\avg{\hat{J}_+^2}-5\avg{\hat{J}_+}^2|\avg{\hat{J}_+}|^2\right)-\chi_4^*\chi_6\avg{\hat{N}_a}\avg{\hat{J}_+^2}\right\}\Bigg)\nonumber \\
&&+\gamma_a\Bigg(\left(\abs{\chi_1}^2+\abs{\chi_3}^2\avg{\hat{J}_-\hat{J}_+}+\abs{\chi_5}^2\avg{\hat{J}_-\hat{J}_+}\right)\avg{\hat{N}_a}+\left(\abs{\chi_2}^2+\abs{\chi_6}^2\avg{\hat{J}_-\hat{J}_+}+\abs{\chi_4}^2\avg{\hat{J}_-\hat{J}_+}\right)\avg{\hat{N}_b}\nonumber \\
&&+2\Re\left\{\chi_1^*\chi_6\avg{\hat{J}_+^2}\right\}+2\Re\left\{\chi_1^*\chi_4\right\}\avg{\hat{J}_+\hat{J}_-}+2\Re\left\{\chi_2^*\chi_3\right\}\avg{\hat{J}_-\hat{J}_+}\nonumber \\
&&+2\Re\left\{\left(\chi_2\chi_5^*+\chi_3\chi_5^*\avg{\hat{N}_a}+\chi_4^*\chi_6\avg{\hat{N}_b}\right)\avg{\hat{J}_+^2}\right\}\Bigg)\\
\frac{d}{dt}\avg{\hat{J}_{-}\hat{J}_+}&=&4\avg{\hat{J}_z}\Im\left\{H_{++}\avg{\hat{J}^2_+}\right\}+2\left(\abs{\kappa_1}^2\avg{\hat{J}_+\hat{J}_-}-\abs{\kappa_2}^2\avg{\hat{J}_-\hat{J}_+}\right)\avg{\hat{J}_z}-\Gamma\Bigg(\left(\abs{\chi_1}^2+\abs{\chi_2}^2\right)\avg{\hat{J}_-\hat{J}_+}\nonumber \\
&&-\abs{\chi_1}^2\hat{N}_b+\left(\abs{\chi_3}^2+\abs{\chi_6}^2\right)\left(\avg{\hat{J}_-}^2\avg{\hat{J}_+^2}+\avg{\hat{J}_+}^2\avg{\hat{J}_-^2}+4|\avg{\hat{J}_+}|^2\avg{\hat{J}_-\hat{J}_+}-5|\avg{\hat{J}_+}|^4\right)+\nonumber \\
&&+2\left(\abs{\chi_3}^2\avg{\hat{N}_a}+\abs{\chi_6}^2\avg{\hat{N}_b}\right)\avg{\hat{J}_z}\avg{\hat{J}_-\hat{J}_+}-\abs{\chi_3}^2\avg{\hat{N}_b}\avg{\hat{J}_-\hat{J}_+}-\abs{\chi_5}^2\avg{\hat{N}_b}\avg{\hat{J}_+\hat{J}_-} \nonumber \\ 
&&+\left(\abs{\chi_5}^2+\abs{\chi_4}^2\right)\left(\avg{\hat{J}_-}^2\avg{\hat{J}_+^2}+\avg{\hat{J}_+}^2\avg{\hat{J}_-^2}+|\avg{\hat{J}_+}|^2\left(3\avg{\hat{J}_+\hat{J}_-}+\avg{\hat{J}_-\hat{J}_+}\right)-5|\avg{\hat{J}_+}|^4\right) \nonumber \\
&&-2\left(\abs{\chi_5}^2\avg{\hat{N}_a}+\abs{\chi_4}^2\avg{\hat{N}_b}\right)\avg{\hat{J}_z}\avg{\hat{J}_+\hat{J}_-}+2\Re\left\{\chi_1^*\chi_6\avg{\hat{J}_+^2}\right\}\left(\avg{N_a}-1\right)\nonumber \\
&&+2\Re\left\{\chi_1^*\chi_4\right\}\left(\avg{\hat{N}_b}-1\right)\avg{\hat{J}_+\hat{J}_-}+2\Re\left\{\chi_2^*\chi_3\right\}\left(\avg{\hat{N}_a}-1\right)\avg{\hat{J}_-\hat{J}_+}+2\Re\left\{\chi_2\chi_5^*\avg{\hat{J}_-^2}\right\}\left(\avg{\hat{N}_b}-1\right)\nonumber \\
&&-2\Re\left\{\left(\chi_4\chi_6^*+\chi_3^*\chi_5\right)\left(\avg{\hat{J}_+}^2\left(2\avg{\hat{J}_-\hat{J}_+}+\avg{\hat{J}_+\hat{J}_-}\right)+3|\avg{\hat{J}_+}|^2\avg{\hat{J}_-^2}-5\avg{\hat{J}_-}^2|\avg{\hat{J}_+}|^2\right)-\chi_3^*\chi_5\avg{\hat{N}_b}\avg{\hat{J}_-^2}\right\}\Bigg)\nonumber \\
&&+\gamma_b\Bigg(\left(\abs{\chi_1}^2+\abs{\chi_3}^2\avg{\hat{J}_-\hat{J}_+}+\abs{\chi_5}^2\avg{\hat{J}_+\hat{J}_-}\right)\avg{\hat{N}_a}+\left(\abs{\chi_2}^2+\abs{\chi_6}^2\avg{\hat{J}_-\hat{J}_+}+\abs{\chi_4}^2\avg{\hat{J}_+\hat{J}_-}\right)\avg{\hat{N}_b}\nonumber \\
&&+2\Re\left\{\chi_1^*\chi_6\avg{\hat{J}_+^2}\right\}+2\Re\left\{\chi_1^*\chi_4\right\}\avg{\hat{J}_+\hat{J}_-}+2\Re\left\{\chi_2^*\chi_3\right\}\avg{\hat{J}_-\hat{J}_+}\nonumber \\
&&+2\Re\left\{\left(\chi_2\chi_5^*+\chi_3\chi_5^*\avg{\hat{N}_a}+\chi_4^*\chi_6\avg{\hat{N}_b}\right)\avg{\hat{J}_+^2}\right\}\Bigg) \\
\frac{d}{dt}\avg{\hat{N}_a}&=&-2\Im\left\{H_{++}\avg{\hat{J}_+^2}\right\}-\abs{\kappa_1^2}\avg{\hat{J}_+\hat{J}_-}+\abs{\kappa_2^2}\avg{\hat{J}_-\hat{J}_+}+\gamma_a\Bigg(\abs{\chi_2}^2\avg{\hat{N}_b}+\left(\abs{\chi_3}^2\avg{\hat{J}_-\hat{J}_+}-\abs{\chi_5}^2\avg{\hat{J}_+\hat{J}_-}\right)\avg{\hat{N}_a}\nonumber \\
&&+2\abs{\chi_6}^2\avg{\hat{N}_b}\avg{\hat{J}_-\hat{J}_+}+2\Re\left\{\chi_1^*\chi_6\avg{\hat{J}_+^2}\right\}+2\Re\left\{\chi_2^*\chi_3\right\}\avg{\hat{J}_-\hat{J}_+}+2\Re\left\{\chi_4^*\chi_6\avg{\hat{J}_+^2}\right\}\avg{\hat{N}_b}\Bigg)-\left(\gamma_b+\gamma_o\right)\Bigg(\nonumber \\
&&\left(\abs{\chi_1}^2+2\abs{\chi_5}^2\avg{\hat{J}_+\hat{J}_-}\right)\avg{\hat{N}_a}+\left(\abs{\chi_4}^2\avg{\hat{J}_+\hat{J}_-}-\abs{\chi_6}^2\avg{\hat{J}_-\hat{J}_+}\right)\avg{\hat{N}_b}+2\Re\left\{\chi_1^*\chi_4\right\}\avg{\hat{J}_+\hat{J}_-}\nonumber \\
&&+2\Re\left\{\chi_2\chi_5^*\avg{\hat{J}_+^2}\right\}+2\Re\left\{\chi_3\chi_5^*\avg{\hat{J}_+^2}\right\}\avg{\hat{N}_a}\Bigg)\\
\frac{d}{dt}\avg{\hat{N}_b}&=&2\Im\left\{H_{++}\avg{\hat{J}_+^2}\right\}+\abs{\kappa_1^2}\avg{\hat{J}_+\hat{J}_-}-\abs{\kappa_2^2}\avg{\hat{J}_-\hat{J}_+}-\left(\gamma_a+\gamma_o\right)\Bigg(\left(\abs{\chi_3}^2\avg{\hat{J}_-\hat{J}_+}-\abs{\chi_5}^2\avg{\hat{J}_+\hat{J}_-}\right)\avg{\hat{N}_a}\nonumber \\
&&+\abs{\chi_2}^2\avg{\hat{N}_b}+2\abs{\chi_6}^2\avg{\hat{N}_b}\avg{\hat{J}_-\hat{J}_+}+2\Re\left\{\chi_1^*\chi_6\avg{\hat{J}_+^2}\right\}+2\Re\left\{\chi_2^*\chi_3\right\}\avg{\hat{J}_-\hat{J}_+}+2\Re\left\{\chi_4^*\chi_6\avg{\hat{J}_+^2}\right\}\avg{\hat{N}_b}\Bigg)\nonumber \\
&&+\gamma_b\Bigg(\left(\abs{\chi_1}^2+2\abs{\chi_5}^2\avg{\hat{J}_+\hat{J}_-}\right)\avg{\hat{N}_a}+\left(\abs{\chi_4}^2\avg{\hat{J}_+\hat{J}_-}-\abs{\chi_6}^2\avg{\hat{J}_-\hat{J}_+}\right)\avg{\hat{N}_b}+2\Re\left\{\chi_1^*\chi_4\right\}\avg{\hat{J}_+\hat{J}_-}\nonumber \\
&&+2\Re\left\{\chi_2\chi_5^*\avg{\hat{J}_+^2}\right\}+2\Re\left\{\chi_3\chi_5^*\avg{\hat{J}_+^2}\right\}\avg{\hat{N}_a}\Bigg) \\
\frac{d}{dt}\avg{\hat{J}_+}&=&2i\avg{\hat{J}_z}\left(\Re\left\{H_{+-}\right\}\avg{\hat{J}_+}+H_{++}^*\avg{\hat{J}_-}\right)+\left(\abs{\kappa_1}^2\left(\avg{\hat{J}_z}-1\right)-\abs{\kappa_2}^2\avg{\hat{J}_z}\right)\avg{\hat{J}_+}+\kappa_1\kappa_2^*\avg{\hat{J}_-}-\Gamma\Bigg(\nonumber \\
&&\frac{\abs{\chi_1}^2+\abs{\chi_2}^2}{2}\avg{\hat{J}_+}+\frac{\abs{\chi_3}^2+\abs{\chi_6}^2}{2}\left(\avg{\hat{J}_-}\avg{\hat{J}_+^2}+2\avg{\hat{J}_+}\avg{\hat{J}_-\hat{J}_+}-2|\avg{\hat{J}_+}|^2\avg{\hat{J}_+}\right)\nonumber \\
&&+\left(\abs{\chi_3}^2\avg{\hat{N}_a}+\abs{\chi_6}^2\avg{\hat{N}_b}\right)\avg{\hat{J}_z}\avg{\hat{J}_+}-\left(\abs{\chi_5}^2\avg{\hat{N}_a}+\abs{\chi_4}^2\avg{\hat{N}_b}\right)\avg{\hat{J}_z}\avg{\hat{J}_+}\nonumber \\
&&+\frac{\abs{\chi_4}^2+\abs{\chi_5}^2}{2}\left(\avg{\hat{J}_-}\avg{\hat{J}_+^2}+2\avg{\hat{J}_+}\avg{\hat{J}_+\hat{J}_-}-2|\avg{\hat{J}_+}|^2\avg{\hat{J}_+}\right)-\chi_1^*\chi_4\avg{\hat{J}_z}\avg{\hat{J}_+}+\chi_1\chi_4^*\left(\avg{\hat{N}_a}+\avg{\hat{N}_b}\right)\avg{\hat{J}_+}\nonumber \\
&&+\left(\chi_2^*\chi_3\frac{\avg{\hat{N}_a}+\avg{\hat{N}_b}}{2}+\chi_2\chi_3^*\avg{\hat{J}_z}\right)\avg{\hat{J}_+}+\chi_2^*\chi_5\avg{\hat{N}_b}\avg{\hat{J}_-}+\frac{\chi_4^*\chi_6+\chi_3\chi_5^*}{2}\left(3\avg{\hat{J}_+^2}-2\avg{\hat{J}_+}^2\right)\avg{\hat{J}_+}\nonumber \\
&&+\frac{\chi_4\chi_6^*+\chi_3^*\chi_5}{2}\left(\left(\avg{\hat{J}_+\hat{J}_-}+\avg{\hat{J}_-\hat{J}_+}\right)\avg{\hat{J}_-}+\avg{\hat{J}_+}\avg{\hat{J}_-^2}-2|\avg{\hat{J}_+}|^2\avg{\hat{J}_-}\right)+\chi_1\chi_6^*\avg{\hat{N}_a}\avg{\hat{J}_-}\Bigg)
\end{eqnarray}
Furthermore, we have that $\frac{d}{dt}\avg{\hat{J}_z}=\frac{1}{2}\left(\frac{d}{dt}\avg{\hat{N}_a}-\frac{d}{dt}\avg{\hat{N}_b}\right),\frac{d}{dt}\avg{\hat{J}_-^2}=\left(\frac{d}{dt}\avg{\hat{J}_+^2}\right)^{\dagger}$, and $\frac{d}{dt}\avg{\hat{J}_-}=\left(\frac{d}{dt}\avg{\hat{J}_+}\right)^{\dagger}$. 
\end{widetext}

\section{Effects of cavity dissipation} \label{app:C}
In this appendix we provide some intuition for the effects of cavity dissipation in the two-axis twisting scheme. With appropriate laser detunings, we can engineer the following effective Hamiltonian and collective Lindblad operator (which appear as Eqs. (\ref{eq:twoeffect}) and (\ref{eq:leff2}) in the main text):

\begin{eqnarray}
    \hat{H}_{\text{eff2}} &=& \frac{i \alpha}{2} \left[ \hat{J}_+^2 - \hat{J}_-^2 \right] \nonumber \\
        &=& \frac{\alpha}{2} \left[ (\hat{J}_x-\hat{J}_y)^2 - (\hat{J}_x+\hat{J}_y)^2 \right] \\
        \nonumber \\
    \hat{L}_{c,\text{eff2}} &=& \sqrt{\frac{\gamma_c}{4}} \left[ (i \hat{J}_++\hat{J}_-) e^{i \delta t + i \phi} - (i \hat{J}_+-\hat{J}_-) e^{-i \delta t - i \phi} \right] \nonumber \\
        &=& \sqrt{\frac{\gamma_c}{2}} \left[ (\hat{J}_x - \hat{J}_y) e^{i \delta t + i \phi + i \pi / 4} \right. \nonumber \\
        && \left. + (\hat{J}_x + \hat{J}_y) e^{-i \delta t - i \phi - i \pi / 4} \right]
\end{eqnarray}
where $\alpha = 16 \left|\chi\right|^2 \delta / (4 \delta^2 + \tilde{\kappa}^2)$, $\gamma_c = 16 \left|\chi\right|^2 \kappa / (4 \delta^2 + \tilde{\kappa})$, and $\phi = \text{Arg} (\delta + i \tilde{\kappa}/2)$. The dynamics of the quantum state under $\hat{H}_{\text{eff2}}$ and $\hat{L}_{c,\text{eff2}}$ are described by the quantum master equation:
\begin{eqnarray}
    \frac{d}{dt}\hat{\rho} &=& -i \left[ \hat{H}_{\text{eff2}}, \hat{\rho} \right] + \hat{L}_{c,\text{eff2}} \ \hat{\rho} \left( \hat{L}_{c,\text{eff2}} \right)^{\dagger} \nonumber \\
    && - \frac{1}{2} \hat{\rho} \left( \hat{L}_{c,\text{eff2}} \right)^{\dagger} \hat{L}_{c,\text{eff2}} \nonumber \\
    && - \frac{1}{2} \left( \hat{L}_{c,\text{eff2}} \right)^{\dagger} \hat{L}_{c,\text{eff2}} \ \hat{\rho}.
\end{eqnarray}\par
The Lindblad operator $\hat{L}_{c,\text{eff2}}$ consists of two terms that accumulate opposite phases $e^{\pm i \delta t}$. If we expand each term in the master equation, we get some terms whose phase factors cancel, and others with phases at $e^{\pm 2 i \delta t}$, e.g.:
\begin{eqnarray}
    \left( \hat{L}_{c,\text{eff2}} \right)^{\dagger} \hat{L}_{c,\text{eff2}} &=& \frac{\gamma_c}{2} \left[ (\hat{J}_x-\hat{J}_y)^2 + (\hat{J}_x+\hat{J}_y)^2 \right.\nonumber \\
    && + (\hat{J}_x+\hat{J}_y)(\hat{J}_x-\hat{J}_y) e^{i 2 \delta t + i 2 \phi + i \pi/2} \nonumber \\
    && \left. + (\hat{J}_x-\hat{J}_y)(\hat{J}_x+\hat{J}_y) e^{-i 2 \delta t - i 2 \phi - i \pi/2} \right]. \nonumber
\end{eqnarray}
Appealing to an argument similar to the rotating wave approximation, we can neglect the pair of rapidly oscillating terms, so long as we are interested in timescales that are long compared to the detuning $\delta$. This approximation decouples the two terms in the original Lindblad operator, and gives a pair of two new Lindblad operators instead:
\begin{eqnarray}
    \hat{L}_{c,\text{eff2}}^{(1)} &=& \sqrt{\frac{\gamma_c}{2}} \left[ \hat{J}_x + \hat{J}_y \right] \nonumber \\
    \hat{L}_{c,\text{eff2}}^{(2)} &=& \sqrt{\frac{\gamma_c}{2}} \left[ \hat{J}_x - \hat{J}_y \right]
\end{eqnarray}\par
This pair of operators generates isotropic spreading of the Wigner function in the $J_x$-$J_y$ plane, as sketched in Fig. (\ref{fig:figure0}b) in the main text. In fact, one can show that the action of this pair of operators is equivalent to any pair of Lindblad operators of the form:
\begin{eqnarray}
    \hat{L}_{c,\text{eff2}}^{(1)} &=& \sqrt{\gamma_c} \left[ \hat{J}_x \cos \theta + \hat{J}_y \sin \theta \right] \nonumber \\
    \hat{L}_{c,\text{eff2}}^{(2)} &=& \sqrt{\gamma_c} \left[ -\hat{J}_x \sin \theta + \hat{J}_y \cos \theta \right]
\end{eqnarray}
which implies that there is no preferred axis for this dissipation.

\end{document}